\documentclass[fleqn,usenatbib]{mnras}

\usepackage{newtxtext,newtxmath}

\usepackage[T1]{fontenc}
\usepackage{ae,aecompl}
\usepackage{ulem} 

\usepackage{appendix}
\usepackage{threeparttable}

\usepackage{graphicx}	
\usepackage{amsmath}	

\title[The cosmic environment holds the key to remain galaxy star formation]{The cosmic environment overtakes the local density in shaping galaxy star formation}

\author[J. Ren et al.]{
Jian~Ren,$^{1,2}$
Zhizheng~Pan,$^{1,2}$\thanks{panzz@pmo.ac.cn}
X.~Z.~Zheng,$^{1,2}$\thanks{ xzzheng@pmo.ac.cn}
Jianbo~Qin,$^{1}$
D.~D.~Shi,$^{1}$
Valentino~Gonzalez,$^{3,4}$
Fuyan~Bian,$^{5}$
\and
Jia-Sheng ~Huang,$^{6,7}$
Min~Fang,$^{1,2}$
Wenhao~Liu,$^{1,2}$
Run~Wen,$^{1,2}$
Yuheng~Zhang,$^{1,2}$
Man~Qiao,$^{1,2}$
and Shuang~Liu$^{1,2}$
\\
$^{1}$Purple Mountain Observatory, Chinese Academy of Sciences, 10 Yuanhua Road, Nanjing 210023, China\\
$^{2}$School of Astronomy and Space Sciences, University of Science and Technology of China, Hefei 230026, China \\
$^{3}$Chinese Academy of Sciences South America Centre for Astronomy, China-Chile Joint Centre for Astronomy, Camino del Observatorio 1515, Las Condes, Chile \\
$^{4}$Centro de Astrofísica y Tecnologías Afines (CATA), Camino del Observatorio 1515, Las Condes, Santiago, Chile \\
$^{5}$European South Observatory, Alonso de Cordova 3107, Casilla 19001, Vitacura, Santiago 19, Chile\\
$^{6}$Chinese Academy of Sciences South America Center for Astronomy, National Astronomical Observatories, Chinese Academy of Sciences, Beijing 100101, China\\
$^{7}$CAS Key Laboratory of Optical Astronomy, National Astronomical Observatories, Chinese Academy of Sciences, Beijing 100101, China
}

\date{Accepted 2021 December 6; Received  2021 December 1; in original form 2021 September 29}

\pubyear{2021}

\begin{document}
\label{firstpage}
\pagerange{\pageref{firstpage}--\pageref{lastpage}}
\maketitle

\begin{abstract}
The gas supply from the cosmic web is the key to sustain star formation in galaxies. It remains to be explored how the cosmic large-scale structure (LSS) effects on galaxy evolution at given local environments. We examine galaxy specific star formation rate as a function of local density in a LSS at $z=0.735$ in the Extended Chandra Deep Field South. The LSS is mapped by 732 galaxies with $R<24$\,mag and redshift at $0.72\le z \le 0.75$ collected from the literature and our spectroscopic observations with Magellan/IMACS, consisting of five galaxy clusters/groups and surrounding filaments over an area of $23.9 \times22.7$\,co-moving\,Mpc$^2$. The spread of spectroscopic redshifts corresponds a velocity dispersion of 494\,km\,s$^{-1}$, indicating the LSS likely to be a thin sheet with a galaxy density $\gtrsim 3.9$ times that of the general field.  These clusters/groups in this LSS mostly exhibit elongated morphologies and multiple components connected with surrounding filaments.  Strikingly, we find that star-forming galaxies in the LSS keep star formation at the same level as field, and show no dependence on local density but stellar mass. Meanwhile, an increasing fraction of quiescent galaxies is detected at increasing local density in both the LSS and the field, consistent with the expectation that galaxy mass and local dense environment hold the key to quench star formation.  Combined together, we conclude that  the cosmic environment of the LSS overtakes the local environment in remaining galaxy star formation to the level of the field.
\end{abstract}


\section{Introduction}

Galaxies which are formed at peaks of initial dark matter density fields \citep[e.g.][]{Davis1985} acquire their angular momentum in the early Universe under a great influence of tidal torque during their formation \citep[e.g.][see \citealp{Schafer2009} for a review]{Peebles1969}. As galaxies and dark matter are not randomly distributed but arranged in the intricate `cosmic web', the large-scale structure (LSS) of the Universe with scales up to tens of Mpc consists of relatively high-density clusters, filaments and relatively low-density voids \citep[e.g.][]{Shandarin1989,Bond1996}. LSSs are ideal laboratories for studying galaxy formation and evolution in different environments. Mapping LSSs and intergalactic medium (IGM) in connection with the properties of galaxies have become a central task in extragalactic astronomy.

Theoretically, galaxies reside in low-mass dark matter haloes (less massive than the critical shock-heating mass of $M_{\rm c}\sim10^{12}$\,M$_\odot$) can accrete cold gas from cosmic web to feed star formation directly. Above $M_{\rm c}$, the accreted gas will be shock-heated to the virial temperature before settling to the halo centre \citep{Keres2005, Dekel2006}.
Since the cooling of the hot gas becomes inefficient possibly due to energy injection from the active galactic nucleus feedback \citep{Somerville2015}, the stellar mass ($M\ast$) growth in massive haloes is suppressed. However, some studies suggest that cold gas can penetrate the hot gas halo to feed star formation through cold flows at high redshift ($z$) \citep{Dekel2006, Keres2009, Ocvirk2008, Almeida2014}. There has been some observational evidence for the existence of such cold flows in high-$z$ protoclusters \citep{Daddi2021}. Investigating the star formation properties of galaxies in filaments and clusters thus may shed light on the gas accretion manners in these environments and how cold flows occur.

The star formation properties of galaxies depend, on the one hand, on the amount of the accreted gas and the star formation efficiency, and on the other hand, on the environment in which the galaxies are located. In low-$z$, many observed properties of galaxies are strongly correlated with their environment \citep[e.g.][]{Gomez2003, Kauffmann2004, Baldry2006}. It is believed that many early-type galaxies  in clusters are formed through galaxy mergers. After that, they sink into the central region of clusters during dynamic relaxation. Late-type galaxies gradually lost their gas reservoirs when they are accreted into the inner region of clusters and evolve into lenticular galaxies (S0). For those galaxies that are far away from clusters, they are less influenced by environmental effects and still have disc or irregular morphologies. This picture can explain the origin of the well known morphology$-$density relation in low redshift\citep[e.g.][]{Dressler1980, Goto2003, Holden2007}.  

Much efforts have been devoted to study the morphology$-$density relation in clusters \citep{Dressler1997, Fasano2000, Postman2005, Smith2005}, star formation activities \citep{Kauffmann2004, Brinchmann2004, Asari2007, Poggianti2008, Peng2010, Ricciardelli2014}  and stellar populations \citep{ Vulcani2012, Guglielmo2018, Guglielmo2019} in different environments  since $z=1$. It is expected that at $z\sim1$, star formation rate (SFR) of cluster member galaxies decreases toward denser environments. However, such a trend is not seen for star-forming member galaxies \citep{Balogh2004, Patel2009, Patel2011, Muzzin2012}. This effect is mainly due to the existence of high proportion of quiescent galaxies (QGs) in dense regions, and the lower SFRs of such galaxies lead to the establishment of the SFR$-$density relation \citep{Scoville2013,Chartab2020}.The environment effects still play a role in shaping galaxy properties at $1<z<2$ \citep{van der Burg2020, Old2020}. At $z>2$, the galaxy star formation properties show no environment dependence \citep{Scoville2013, Darvish2016}.  Observationally, the cosmic star formation density declines rapidly since $z\sim2$ \citep{Madau2014} along with a significant growth of the  QG population \citep{Muzzin2013,Buitrago2013}. This is likely a consequence of the decrease of gas accretion onto galaxies since the cosmic star formation peak \citep{Pan2019}. 

The local and large-scale environment influence galaxy evolution with very different manners. In a dense local environment, frequent galaxy interactions can lead to the formation of disturbed morphologies or tidal tails \citep{Ren2020}. In the extreme cases, the stripped gas and stars from interacting galaxies can form ultra-compact galaxies or ultra-diffuse galaxies \citep{Pfeffer2013, Janssens2017}. In addition, interactions between central and satellite galaxies in groups can cause gas removal, which can suppress the star formation of satellite galaxies \citep{van de Voort2017,Kawinwanichakij2016}.  In contrast, the large-scale environment influences galaxy evolution via replenish, cut off or rapidly deplete the cold gas supply in galaxies \citep{Dekel2009, Castignani2021}. In the cluster environment, it is well acknowledged that the hot intergalactic medium will remove gas from galaxies through ram pressure stripping \citep[e.g.][]{Bekki2014,Bahe2015,Hester2006}. In the filament environment, the observations have not yet converged to the same conclusion. For example, \cite{Kleiner2017} find evidence for the replenishment of H\,{\footnotesize I} gas in massive galaxies from the intra-filament medium in the low-$z$ Universe.
However, more studies support a picture where galaxies in filaments are cut off from their gas supply relative to those in voids \citep{Kuutma2017,Poudel2017,Crone2018}. Recently, simulation works based on IllustrisTNG-100 suggest that the influence exerted by the large-scale galaxy environment on the circumgalactic medium gas angular momentum results in either enhanced or suppressed or star formation inside a galaxy \citep{Lu2021,Wang2021}. A comparison study for the properties of galaxies in LSSs and general fields is thus important for understanding how the local and large-scale environment influence galaxy evolution.

Over the past two decades, thanks to the Sloan Digital Sky Survey \citep[SDSS;][]{York2000, Stoughton2002, Abazajian2003, Abazajian2004},  massive photometric and spectroscopic data have been available for studying LSSs in the low-$z$ Universe. At intermediate- and high-$z$, the formation and evolution of LSSs and their member galaxies are far less explored. On the one hand, there are few large-area multi-wavelength sky surveys at the intermediate- and high-$z$. On the other hand, accurate redshifts needed for the identification of member galaxies in LSS are also lacking for a large sample of galaxies beyond the local Universe. Many researches have focused on the Extended Chandra Deep Field South (ECDF-S), which has increased spectroscopic redshift (spec-$z$) surveys and multi-wavelength data covering from X-ray \citep{Giacconi2002,Luo2008,Xue2011,Luo2017}, optical to the infrared \citep{Wolf2004, Rix2004, Gawiser2006, Cardamone2010}. These wealthy data make it possible to accurately identify LSSs and study the properties of their member galaxies. In recent years, LSSs and their substructures (e.g. groups, clusters and filaments) in the ECDF-S field have been reported at various redshifts, either from X-ray \citep{Gilli2003} or from optical studies \citep{Adami2005,  Trevese2007, Salimbeni2009, Dehghan2014}.

The LSS at $z=0.735$ in the ECDF-S field has been examined in previous works \citep{Gilli2003,Adami2005,Dehghan2014}. In this work, we aim to investigate relations between galaxy properties and their environment in this LSS. This paper is organized as follows. Section~\ref{sec:sample} introduces our observations and the sample selection of member galaxies in the LSS. Section~\ref{sec:Results} presents the results. A brief discussion is presented in Section~\ref{sec:Discussion}. Section~\ref{sec:Summary} summarizes the conclusion of this work. Throughout this paper, we adopt a concordance cosmology of $H_{0}=70$\,km\,s$^{-1}$\,Mpc$^{-1}$, $\Omega_{\rm m}=0.3$ and $\Omega_{\Lambda}=0.7$. All photometric magnitudes are given in the AB system.

\section{Observations and Sample Selection} \label{sec:sample}

In order to derive a complete member galaxy sample of the $z=0.735$ LSS, all publicly-available spec-$z$ and photometric redshift (phot-$z$)  data in the ECDF-S field are collected. We also carried out new spectroscopic observations using the multi-object spectrograph (MOS) mounted on the Magellan telescope to enlarge the spec-$z$ sample.

\subsection{Public data} \label{sec:Data}

The COMBO-17 survey (Classifying Objects by Medium-Band Observations in 17 filters), centred in the ECDF-S field, covers a field  area of $31\farcm5\times30\arcmin$ with 17 bands ranging from 3\,500 to 9\,300\,\AA. The COMBO-17 catalog contains a total of 63\,501 objects. For the $R<24$\,mag galaxies, the completeness reaches $\sim 80$\,per\,cent and the phot-$z$ accuracy is  $\Delta z /(1+z) \sim 0.06$ \citep{Wolf2004}.
In order to select the sources in the COMBO-17 catalog that have preexisting spec-$z$, we cross-matched the catalog with the published spec-$z$ catalogs in the ECDF-S field. These include the MUltiwavelength Survey by Yale-Chile (MUSYC) catalog complied by \citet{Cardamone2010}, in which they collected all the preexisting spec-$z$ data from \citet{Cristiani2000, Croom2001, Cimatti2002, Le Fevre2004, Strolger2004, Szokoly2004, van der Wel2004,  Kriek2008,  Vanzella2008, Treister2009,  Balestra2010}.  We also use the Arizona CDF-S Environment Survey (ACES) catalog compiled by \citet{Cooper2012}, which is based on a spec-$z$ survey of the Chandra Deep Field South field with Magellan/IMACS. In the recent years, \textit{HST}/WFC3 grism spec-$z$ data \citep{Morris2015} and MUSE wide survey catalog \citep{Urrutia2019} in the Great Observatories Origins Deep Survey South field (GOODS-S)  have also been published. The details of these catalogs are listed in Table~\ref{tab:published data}. We cross-matched the COMBO-17 catalog with these spec-$z$ catalog using an aperture of radius$=1\arcsec$ and generated an ECDF-S Master spec-/phot-Redshift Catalog (EMRC). The EMRC contains 62\,448 sources with $0<z<4.9$,  and 8\,370 of them have spec-$z$. Note that the MUSYC catalog of the ECDF-S field \citep{Cardamone2010} also provides phot-$z$ estimates.  At $0.1<z<1.2$, the phot-$z$ accuracy of the MUSYC catalog is $\Delta z/(1+z) \sim 0.007$, which is better than that of the COMBO-17 catalog. Thus, the MUSYC phot-$z$ will be adopted if a source is matched between the EMRC and MUSYC catalog.

\begin{table}
  \caption{The public redshift catalogs and imaging data in the ECDF-S field.}
  \label{tab:published data}
  \begin{threeparttable}
  \centering
  \begin{tabular}{lcccccc}
    \hline\hline
Name &Telescope& Filed & Type & N\tnote{1} \\
       &  & &  &\\
\hline
COMBO17\tnote{2} &MPG &ECDF-S & phot  & $\sim$\,60\,000 \\ 
GEMS\tnote{3}  & \textit{HST} &CDF-S   & morphology & $\sim$\,10\,000 \\
MUSYC\tnote{4} &Subaru &ECDF-S  & phot  & $\sim$\,40\,000 \\
ACES\tnote{5}  &Magellan &ECDF-S    & spec  & $\sim$\,5\,000 \\ 
WGELRC\tnote{6} & \textit{HST}& GOODS-S  & spec   & $\sim$\,600 \\
MWS\tnote{7} & VLT&CANDELS & spec  &  $\sim$\,2\,000  \\
\hline
 \end{tabular}
 \begin{tablenotes}
        \footnotesize
        \item[1] Number of sources.
        \item[2] Classifying Objects by Medium-Band Observations in 17 Filters \citep{Wolf2004}.
        \item[3] Galaxy Evolution from Morphologies and SEDs \citep{Rix2004}.
        \item[4] MUltiwavelength Survey by Yale-Chile \citep{Cardamone2010}.
        \item[5] Arizona CDF-S Environment Survey (ACES) \citep{Cooper2012}.
        \item[6] WFC3 Grism Emission Line Redshift Catalog \citep{Morris2015}.
        \item[7] MUSE-Wide Survey  \citep{Urrutia2019}.
   \end{tablenotes}
  \end{threeparttable}
 \end{table}

The MUSYC catalog of the ECDF-S field provides 18 medium-bands deep optical photometry from the Subaru telescope and 14 pre-existing broad-band photometry from the $U38$ to \textit{Spitzer}/IRAC 8\,$\micron$. This catalog contains $\sim$\,40\,000 galaxies with $R_{\rm AB}<25.3$\,mag, the median depth of the medium-band imaging.
We use an aperture of $r=0\farcs6$ to cross-match our EMRC with the MUSYC catalog. This aperture radius is equivalent to the coordinate $3\sigma$ deviation of the two catalogs. Finally, we obtain the multi-wavelength fluxes of a sample of 1\,402  galaxies with $R<24$\,mag at $0.69 \le z \le 0.78$. In the following, we will perform spectral energy distribution (SED) fitting to these galaxies to derive their physical properties.

The GOODS-S covering the central region of the ECDF-S field \citep{Giavalisco2004} and the Galaxy Evolution from Morphology and SEDs survey \citep{Rix2004} provide high-resolution imaging data taken with the \textit{Hubble Space Telescope (HST)}, which enables the study of the morphology$-$density relation of member galaxies in the $z\sim0.735$ LSS.

\begin{figure}
	\centering
	\includegraphics[width=\columnwidth]{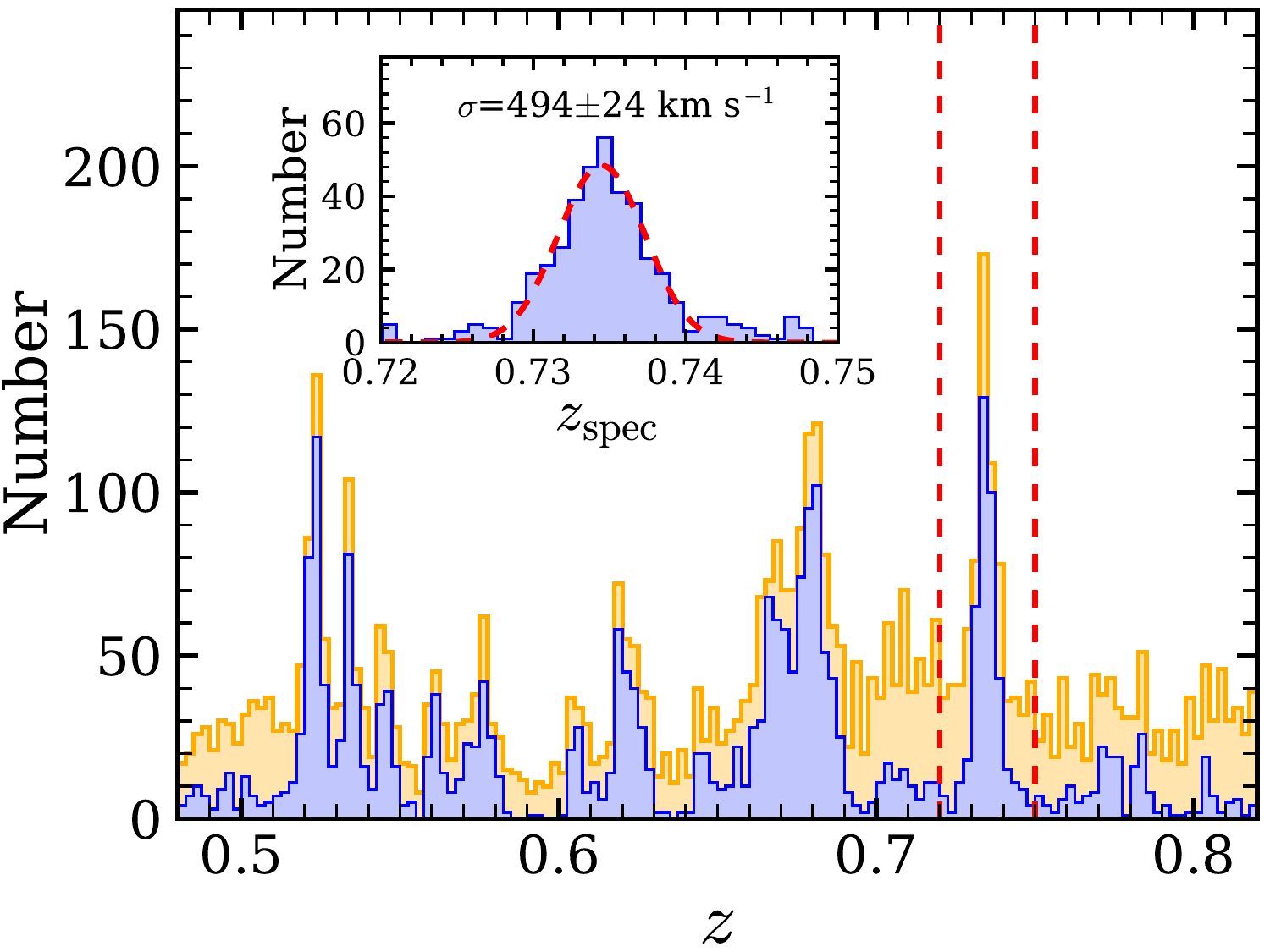}
         \caption{Redshift distribution of $R<24$ mag galaxies in the ECDF-S field, with a bin size $\Delta z=0.0025$. The blue histogram refers to the spectroscopic redshifts, and the orange histogram is for the spec-$z$+phot-$z$. We select the galaxies with $0.72 \le z \le 0.75$ as the members of the LSS. The inner panel shows the spec-$z$ distribution of  the galaxies with a bin size $\Delta z=0.001$ in the LSS. The red dashed curve in the inner panel represents a Gaussian fit to the spec-$z$ distribution.}
         \label{fig:01}
\end{figure}
\subsection{Magellan/IMACS spectroscopic observations} \label{sec:observation}

In order to enlarge the spec-$z$ sample, 125 galaxies with phot-$z$ of $0.7<z<0.8$ are selected from EMRC for spectroscopic observations. We conducted MOS observations on 2018 January 21, using the Inamori-Magellan Areal Camera and Spectrograph \citep[IMACS,][]{Dressler2011} on the Magellan Baade telescope at Las Campanas Observatory. We used the $f/2$ (short) camera on the IMACS with a 200\,line\,mm$^{-1}$ grism (2.037\,\AA \,pixel$^{-1}$) and a broad $r$-band filter with a wavelength coverage of 540$-$700\,nm to do the observations. The total exposure time is 1.5\,h, which is divided into three 1\,800\,s exposures. The median seeing is $0\farcs8$.

The MOS data are reduced using the Carnegie Observatories System for Multi-Object Spectroscopy \footnote{ \url{https://code.obs.carnegiescience.edu/cosmos}} (\texttt{COSMOS}) pipeline \citep{Dressler2011, Oemler2017}.  The pipeline is mainly composed of two major parts: constructing the spectral map and reducing the spectra. In the constructing spectral map step, we subtract bias from each science exposure frame and use the HeNeAr arc to do the wavelength calibration. Then, we trace each spectral position on the science frames. In the reducing spectrum process, we do flat-fielding and sky subtraction from science frames. After that, we extract 2D spectra and stack three exposures to remove cosmic rays. The 1D spectra for all sources are extracted from the reduced 2D spectra using a boxcar extraction, with the extraction aperture matched to the spatial profile of the object.

Considering the spectral wavelength coverage and the redshifts of our target objects, the reduced spectra should cover the rest-frame wavelength range of 3\,000$-$4\,000\,\AA.  There is only one strong emission line [O\,{\footnotesize{II}}]~$\lambda$3727 over the rest-frame wavelength of 3\,000  and 4\,000\,\AA\  in a galaxy spectrum, making a robust redshift identification difficult. Although the [O\,{\footnotesize{II}}]~$\lambda$3727 emission line has doublet, our spectral resolution is not enough to resolve two lines.
The phot-$z$ accuracy of MUSYC is about $\Delta z=0.012$ at $z=0.73$. Therefore, if an emission line is identified between (phot-$z$$-$3$\times$$\Delta z$)$\times 3727$\,\AA\  and (phot-$z$+3$\times$$\Delta z$)$\times 3727$\,\AA\ , we consider it as an  [O\,{\footnotesize{II}}]~$\lambda$3727 emission line. Then, we use \texttt{specutils}\footnote{ \url{https://specutils.readthedocs.io/en/stable/}} to determine the center of the emission line and calculate the spec-$z$.

As such, the spec-$z$ uncertainty $\delta z$ depends on that of the wavelength calibration.  At $z\sim 0.735$, our wavelength calibration results in a redshift uncertainty of $\delta z<0.001$.
This accuracy is sufficiently high for identifying member galaxies in the LSS. Finally, we obtain the spec-$z$ of 75 galaxies from our observations. Examples of 1D and 2D spectra and the catalog of newly observed spec-$z$ from our observations are presented in Appendix~\ref{sec:appendix}.

\begin{figure*}
	\centering
	\includegraphics[width=0.95\textwidth]{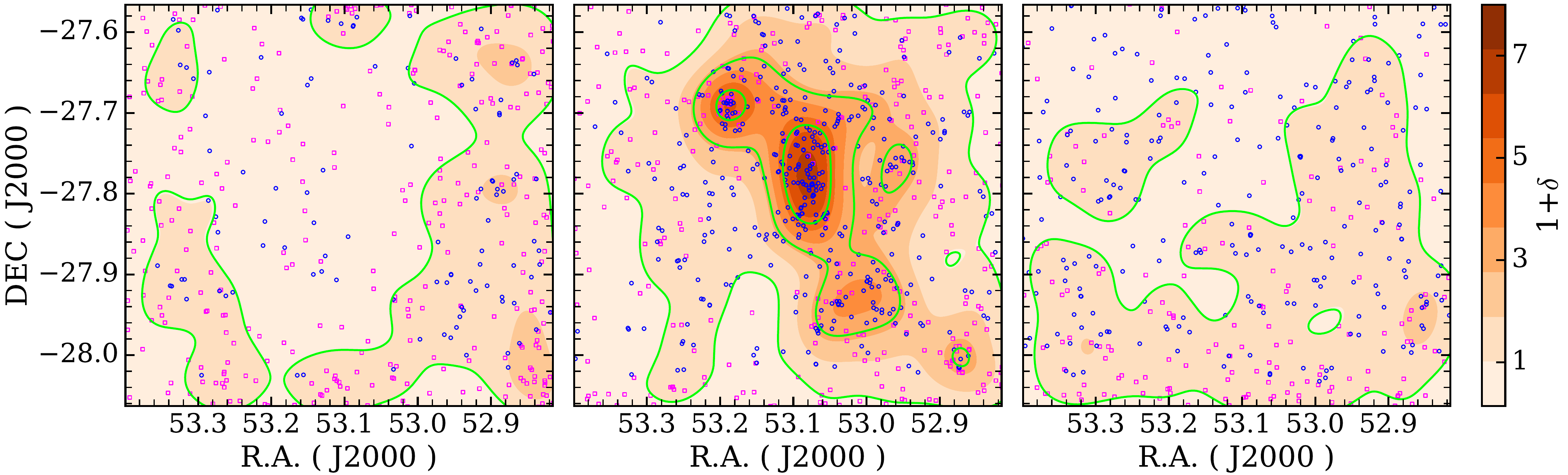}
         \caption{The spatial distribution of galaxies with $R<24$\,mag in the ECDF-S field.  From left to right are the galaxies at $0.69 \le z<0.72$, $0.72 \le z \le 0.75$ and $0.75<z \le 0.78$.  The LSS is shown in the second panel. Blue circles represent the spec-$z$ galaxies and magenta squares represent the phot-$z$ galaxies. The color map shows the relative surface density of all galaxies smoothed by a Gaussian kernel with $\rm{FWHM}=4\farcm6$ (3.5\,cMpc at $z=0.735$). The lime contour levels refer to [1,3,5] times the averages density in the field. The three maps have a size of $31\farcm5\times30\arcmin$, corresponding to a co-moving size of $23.9\times 22.7$\,cMpc$^2$ at $z=0.735$.}
         \label{fig:02}
\end{figure*}

\subsection{Sample selection} \label{sec:Sample selection}
Redshift distribution of galaxies with $R<24$\,mag in the ECDF-S field is shown in Figure~\ref{fig:01}. The golden and blue histograms indicate the spec-$z$+phot-$z$ sample and the spec-$z$ sample, respectively. The inner panel shows the spec-$z$ distribution of the $z=0.735$ LSS. It can be seen that the redshift distribution of the LSS is well fit by a Gaussian function with $\delta z=0.0029$ and $z_{\rm peak}=0.7345$.  Considering the relatively low spec-$z$ completeness and the uncertainty of phot-$z$, we thus select galaxies within the range of $0.72 \le z \le 0.75$ (hereafter the LSS sample) as the member galaxies of the $z=0.735$ LSS. Meanwhile, we selected galaxies with $0.69\le z <0.72$ and $0.75<z \le 0.78$ as a comparison sample (hereafter the field sample), to study the difference in galaxy properties between the LSS and the general field. Finally, the LSS sample contains 732 (414 spec-$z$+318 phot-$z$) galaxies with $R<24$\,mag at $0.72\le z \le 0.75$, and the field sample includes 860 galaxies with $R<24$\,mag at $0.69\le z<0.72$ and $0.75<z \le 0.78$.

\section{Analysis and Results } \label{sec:Results}
Our goal is to study the physical properties of the member galaxies of the $z=0.735$ LSS, and investigate whether these properties are different from those of galaxies reside outside the LSS. We firstly characterize the LSS and its substructures and then study galaxy properties as functions of local environment.

\begin{figure*}
	\centering
	\includegraphics[width=0.8\textwidth]{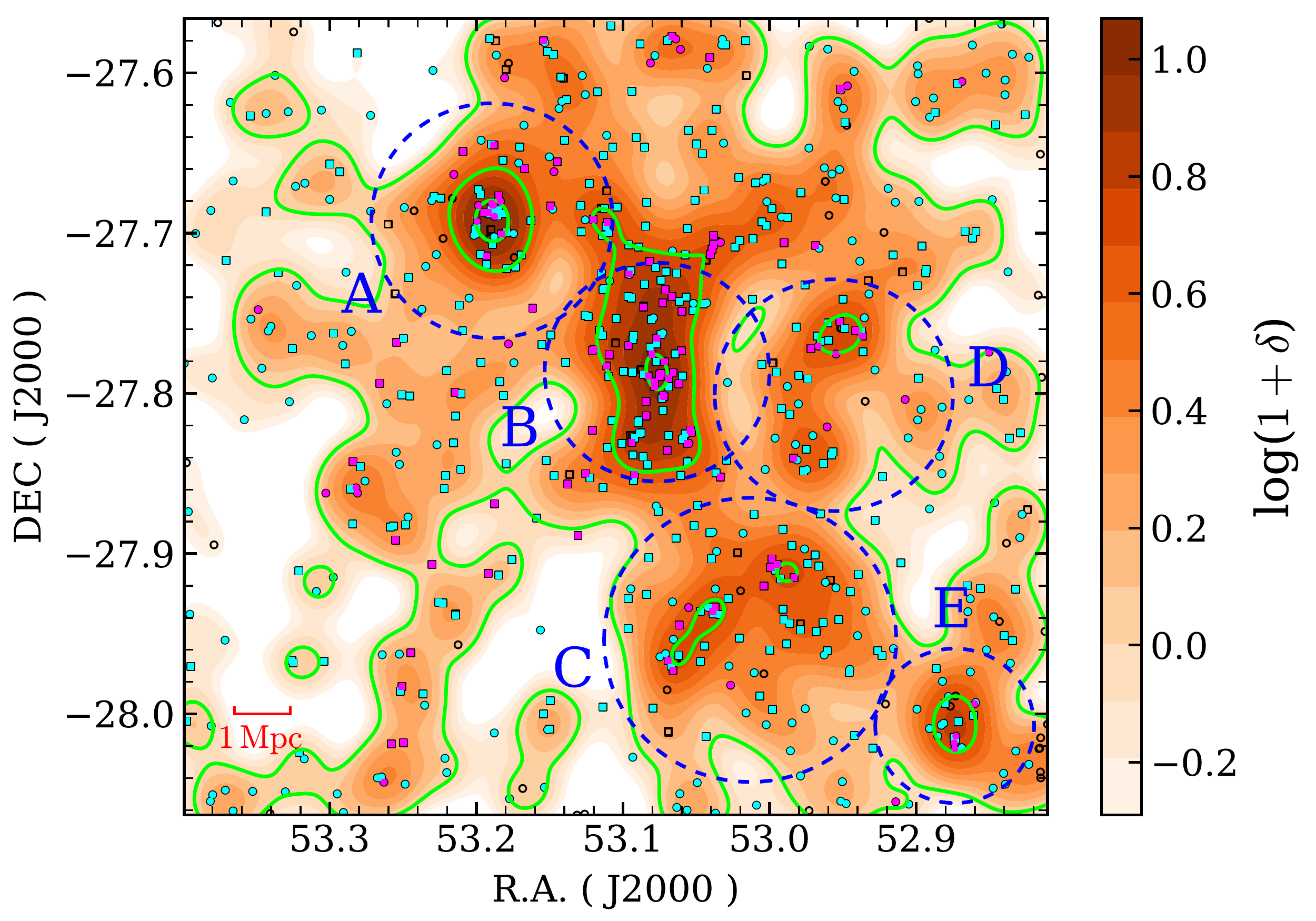}
         \caption{Spatial distribution of member galaxies of the $z = 0.735$ LSS. The squares represent spec-$z$ samples and circles are phot-$z$ samples. Magenta and cyan symbols are QGs and SFGs as divided in Section \ref{sec:SED}. Large blue circles are the $r/R_{200}=3$ region sof the massive substructures detected in the LSS. Unfilled symbols represent no SED-fitting results galaxies or $\chi_r^2>5$ galaxies. Color map shows a density map of $\delta$ relative to the average density in the field. The map is smoothed with a Gaussian Kernel of ${\rm FWHM}=2\farcm3$ (1.75\,cMpc at $z=0.735$).  $1+\delta =1, 5, 10$ are marked with the lime contours.}
         \label{fig:03}
\end{figure*}

\subsection{Spatial distribution of the $z=0.735$ LSS}  \label{sec:Spatial Distribution of The LSS}
 Figure~\ref{fig:02} shows the spatial distribution of LSS member galaxies at $0.72 \le z \le 0.75$ and field galaxies at $0.69\le z<0.72$ and $0.75<z \le 0.78$. Our sample galaxies span a sky area of $31\farcm5 \times 30\arcmin$, corresponding to scales of $23.9\times 22.7$\,co-moving\,Mpc$^2$ (cMpc$^2$) at $z=0.735$.  Blue circles and magenta squares mark spec-$z$ and phot-$z$, respectively.  We split the projected plane into a grid of identical cells with a size of $30\arcsec \times 30\arcsec$, calculate galaxy surface number density in these cells, and smooth the density maps with a Gaussian kernel of $\rm{FWHM}=4\farcm6$ (3.5\,cMpc at $z=0.735$). The FWHM value is larger than the virial radius of general clusters and is suitable for showing 2D density map of the LSS.  The average volume of the $0.69\le z<0.72$ and $0.75<z \le 0.78$ redshift bins are equal to that of the $0.72 \le z \le 0.75$ bin and each of the redshift bins contain nearly the same volume (within  5\,per\,cent).
The surface number densities of the three slices can be compared with each other. The density maps are normalized to the average density in  the field following  $1+\delta=\Sigma/<\Sigma>_{\rm field}$. Here $<\Sigma>_{\rm field}$ is the average density over the two redshift slices $0.69\le z<0.72$ and $0.75<z \le 0.78$.

As shown in the inner panel of Figure~\ref{fig:01}, the spec-$z$ distribution of the LSS member galaxies suggests a scatter of $\delta z=0.0029$. This scatter is a combination of galaxy velocity dispersion  and the cosmic distance along the line-of-sight. Considering that the velocity dispersion of a typical galaxy group reaches $\sim 500$\,km\,s$^{-1}$, and the radial distance of the LSS might be dramatically smaller, implying that this LSS may possibly be a thin sheet. When taking $\pm1.2\times \delta z$, corresponding to the $\Delta z$ equal to the FWHM of the spec-$z$ distribution, as the upper limit for the line-of-sight span of the $z=0.735$ LSS (corresponding to 19.8\,cMpc),  the number density of the LSS  becomes $\gtrsim 3.9$ times the average number density of the field. The number density of the LSS might be significantly higher if the actual size is adopted.

\subsection{Substructures in the $z=0.735$ LSS} \label{sec:Substructures in the LSS}

We further examine the substructures of the $z=0.735$ LSS. It is clear that galaxies within the LSS are concentrated around some density peaks and form substructures including galaxy clusters, massive groups and filaments. These substructures have been identified as small galaxy clusters \citep{Dehghan2014}. Using the latest data we collect, we map the densities in the redshift slice of $0.72 \le z \le 0.75$  and identify five galaxy clusters and filaments around them.

\begin{figure}
	\centering
	\includegraphics[width=\columnwidth]{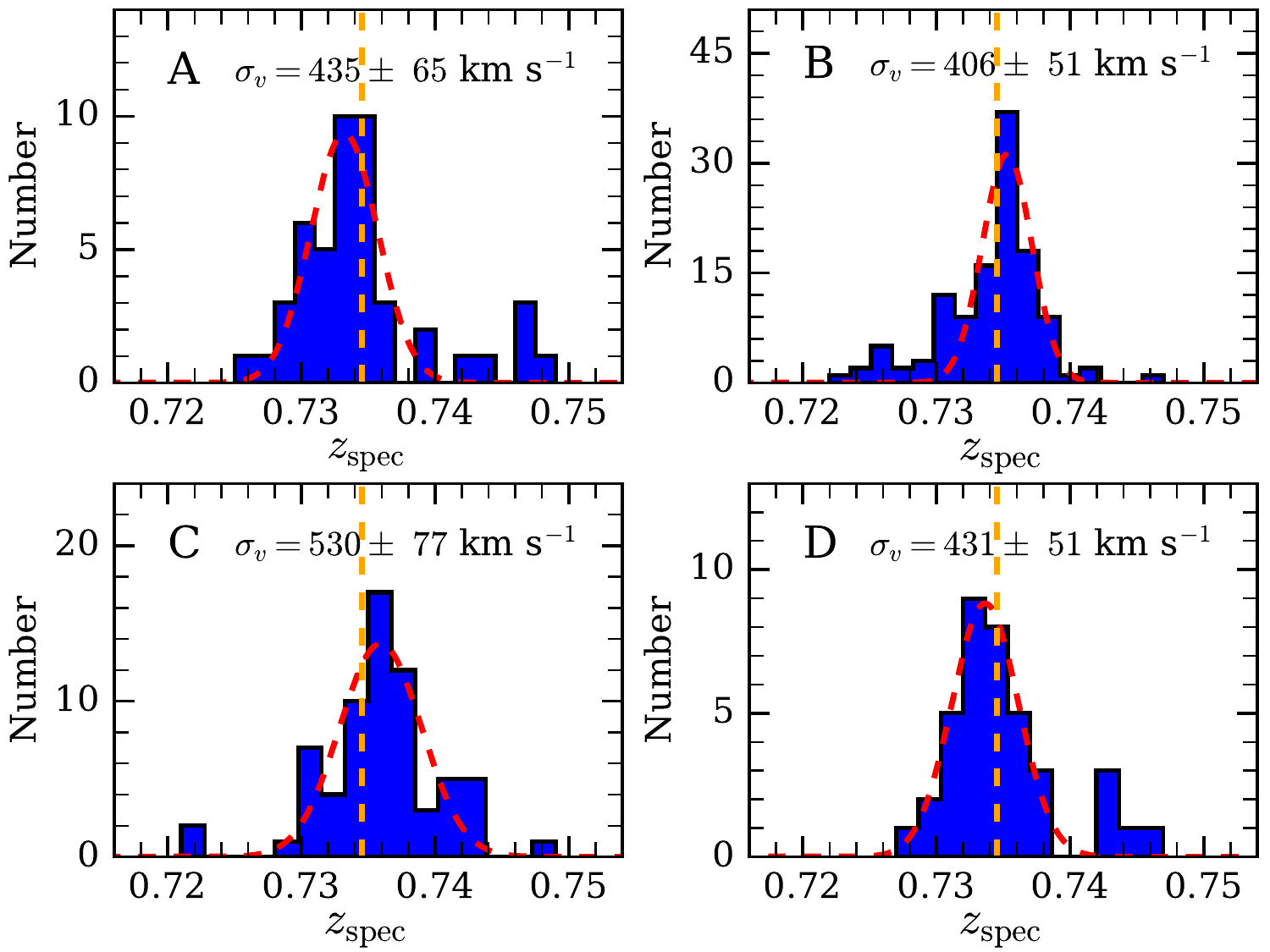}
         \caption{ The spec-$z$ distribution of four clusters. From top-left to bottom-right panels are cluster-A, B, C and D. The red dashed curves show the best-fitting Gaussian functions. The orange dashed lines mark the centre redshift of the LSS. Velocity dispersions of the substructures are given. }
         \label{fig:04}
\end{figure}

In Figure~\ref{fig:03} we mark the five galaxy clusters on the density map of the $z=0.735$ LSS,  smoothed with a Gaussian kernel of $\rm{FWHM}=2\farcm3$ (1.75\,cMpc).  The central density of these substructures must be greater than twice the average density and FWHM$>$1.75\,cMpc. The substructures of the LSS are detected using the one-dimensional density distribution from the centre of the substructures to the first inflection point. Among them, Cluster-A and Cluster-B are galaxy clusters with a single dense core.  Cluster-C and Cluster-D both consist of two smaller substructures. These substructures are common in low-$z$ clusters \citep[e.g.][]{Aguerri2010,Soares2019}. We speculate that the substructures in Cluster-C and Cluster-D will eventually merge and form two massive clusters. Cluster-E has only a few galaxies with spec-$z$. It is thus difficult to robustly determine its velocity dispersion and select member galaxies. The redshift distribution of the member galaxies of four clusters (A, B, C and D) are shown in Figure~\ref{fig:04}.
Note that uncertainties exist with our identification of these substructures as criteria for galaxy  groups, poor clusters and rich clusters are not clearly defined, particularly for unvirialized systems. We remind that our analyses do not rely on the identification of these substructures. Instead, we introduce the density indicator to quantify the local environment of galaxies in Section~\ref{sec:Surface density and galaxies properties}.

\subsubsection{Properties of clusters}  \label{sec:Properties of clusters}
We  fit the redshift distribution of galaxies in each cluster with Gaussian function and obtain the centre redshift ($z_{\rm c}$) and dispersion $\sigma_z$ from the best-fitting function. The observed velocity dispersion ($\sigma_v$) is calculated using $c\sigma_z/(1+z_{\rm c})$, where $c$ is the speed of light. For each galaxy cluster, we estimate $R_{200}$, the radius within which the mass density is 200 times the critical density of the Universe, using the formula from \citet{Carlberg1997} as
\begin{equation}
         R_{200} =\frac{\sqrt 3}{ 10} \frac{\sigma_v}{ H(z_{\rm c})} =\frac{\sqrt 3}{ 10} \frac{\sigma_v}{H_0 \sqrt{\Omega_{\rm m} (1+z_{\rm c})^3 + \Omega_{\Lambda}}}.
\end{equation}
 Assuming that all of the clusters are virialized system, we estimate the mass ($M_{200}$) using the formula from \citet{Voit2005} as

\begin{equation}
         M_{200} = \frac{10^{15}h^{-1}\,M_{\odot}}{ \sqrt{\Omega_{\rm m} (1+z_{\rm c})^3 + \Omega_{\Lambda}}} (\frac{\sigma_v}{1080\, \rm{km\,s^{-1}}})^3.
 \end{equation}
We selected the galaxies within a circular region of  $r/R_{\rm 200}<2$ and $|{z-z_{\rm c}}|< 3\sigma_z $ as the member galaxies of each cluster \citep{Oh2018}. The centre of each region is the smoothed density weighted centre, and the properties of these clusters are listed in Table~\ref{tab:clusters properties}. We also calculate the spec-$z$ completeness (C=N$_{{\rm spec}-z}$/N$_{\rm total}$) within $r/R_{\rm 200}<2$ for each cluster. Clusters that with high spec-$z$ completeness are expected to have more reliable centre redshift and velocity dispersion estimate.

Because these substructures have not relaxed and have higher redshifts, the low-$z$ cluster criterions are not applicable to these substructures. The velocity dispersions, $M_{200}$ and $R_{200}$ of the substructures in our work are similar to that of the low-$z$ small clusters \citep{Soares2019}. The mass and velocity dispersion of the clusters shown in our work are comparable to the Fornax cluster \citep{Drinkwater2001}. Our clusters have higher redshift than the Fornax cluster. We hypothesize that if these clusters evolved from $z =0.735$ to $z=0$, they would be larger than the Fornax cluster.

\begin{table}
\centering
  \caption{Properties of four clusters in the LSS.}
  \label{tab:clusters properties}
   \begin{threeparttable}
  \begin{tabular}{lcccccc}
    \hline\hline
ID & $\sigma_{v}$& $R_{200}$  & M$_{200}$ & N$_{\rm gal}$$\tnote{a}$ & C$\tnote{b}$\\
          & km s$^{-1}$&(Mpc)  &($\times 10^{13}$M$_\odot$)& &\\
\hline

A &   435\,$\pm $\,65 & 0.72\,$\pm $\,0.11 & 6.2$^{+3.2}_{-2.4}$ & 58 & 0.569\\
\\
B &  406\,$\pm $\,51 & 0.67\,$\pm $\,0.08 & 5.0$^{+2.1}_{-1.7}$ & 76 & 0.934\\
\\
C &  530\,$\pm $\,77& 0.87\,$\pm $\,0.13 & 11.2$^{+5.6}_{-4.2}$ & 64 & 0.688\\
\\
D &  431\,$\pm $\,51& 0.71\,$\pm $\,0.08 & 6.0$^{+2.4}_{-1.9}$ & 42  &  0.595\\

\hline
 \end{tabular}
 \begin{tablenotes}
        \footnotesize
        \item[a] Number of member galaxies with spec-$z$+phot-$z$ within $r/R_{200}<2$.
        \item[b] Spec-$z$ completeness within $r/R_{200}<2$ .
 \end{tablenotes}
  \end{threeparttable}
\end{table}

\begin{table}
\centering
  \caption{SED-fitting modules and parameters.}
  \label{tab:sed}
  \begin{tabular}{lccc}
    \hline\hline
Model & Parameters & Value\\
\hline

sfhdelayed &   $\tau$  & 0.05,0.1,0.5,1,2,\\&&4,6,8,10,20 (Gyr) \\
  & age & 1,2,5,7 (Gyr) \\
 {}&$\tau_{\rm{burst}}$ & 5,10,50,100,500 (Myr) \\
 {}& Age$_{\rm{burst}}$ & 10,50,100,200,500 (Myr) \\
 {}& $f_{\rm{burst}}$ & 0.001,0.01,0.02,0.05,0.1 \\
\hline

BC03 & IMF & 1 (Chabrier)   \\
& Metallicity & 0.02\\
\hline

Nebular & Ionisation & $-$2\\
\hline

Calzitti2000 & \it{E(B$-$V)$_{\rm nebular}$} & 0.01,0.05,0.1,0.2,0.3\\&&0.4,0.5,0.6,0.8,1.0,1.2 \\
{}& \it{E(B$-$V)$_{\rm factor}$} & 0.44 \\
{}&  powerlaw\_slope & $-$0.7,$-$0.5,$-$0.3,$-$0.1,0.1,0.3\\
\hline
 \end{tabular}
\end{table}

\subsection{SED fitting} \label{sec:SED}

\begin{figure*}
	\centering
	\includegraphics[width=0.9\textwidth]{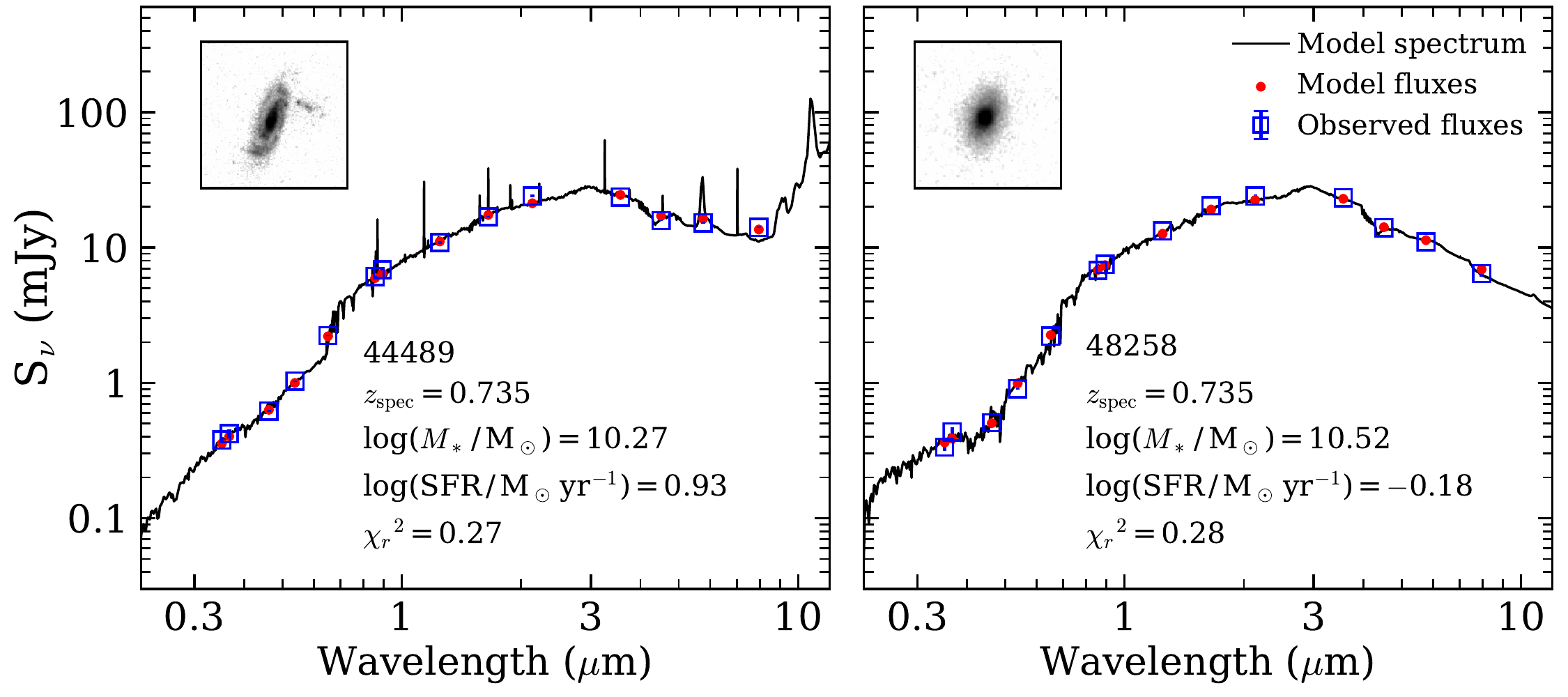}
         \caption{Two Examples of observed galaxy SEDs and best-fitting models. Blue squares with error bars are observed data points and solid lines represent the best-fitting model spectrum. Red dots are the best-fitting fluxes. The inner small panels show the galaxy stamps taken from the \textit{HST}/ACS F775W observations.}
         \label{fig:05}
\end{figure*}

We obtain the broad-band spectral energy distributions (SED) composed of fluxes in the $U38$, $U$, $B$, $V$, $R$, $I$, $z$, $J$, $H$, $K$, 3.6\,$\micron$, 4.5\,$\micron$, 5.8\,$\micron$ and 8.0\,$\micron$ bands for the galaxies with multi-wavelength fluxes at $0.69\le z \le 0.78$ and fit the observed fluxes using a python Code Investigating GALaxy Emission \footnote{https://cigale.lam.fr/} \citep[\texttt{CIGALE};][]{Boquien2019}. Based on the energy balance principle, \texttt{CIGALE} treats the dust attenuation of galaxies, star formation history (SFH), single stellar  population model and other modules as free parameters to fit the observed data and find the best-fitting SED model. A Bayesian analysis is used for constructing a probability distribution function (PDF) of the estimated parameter and then output the adopted values.

\begin{figure}
	\centering
	\includegraphics[width=\columnwidth]{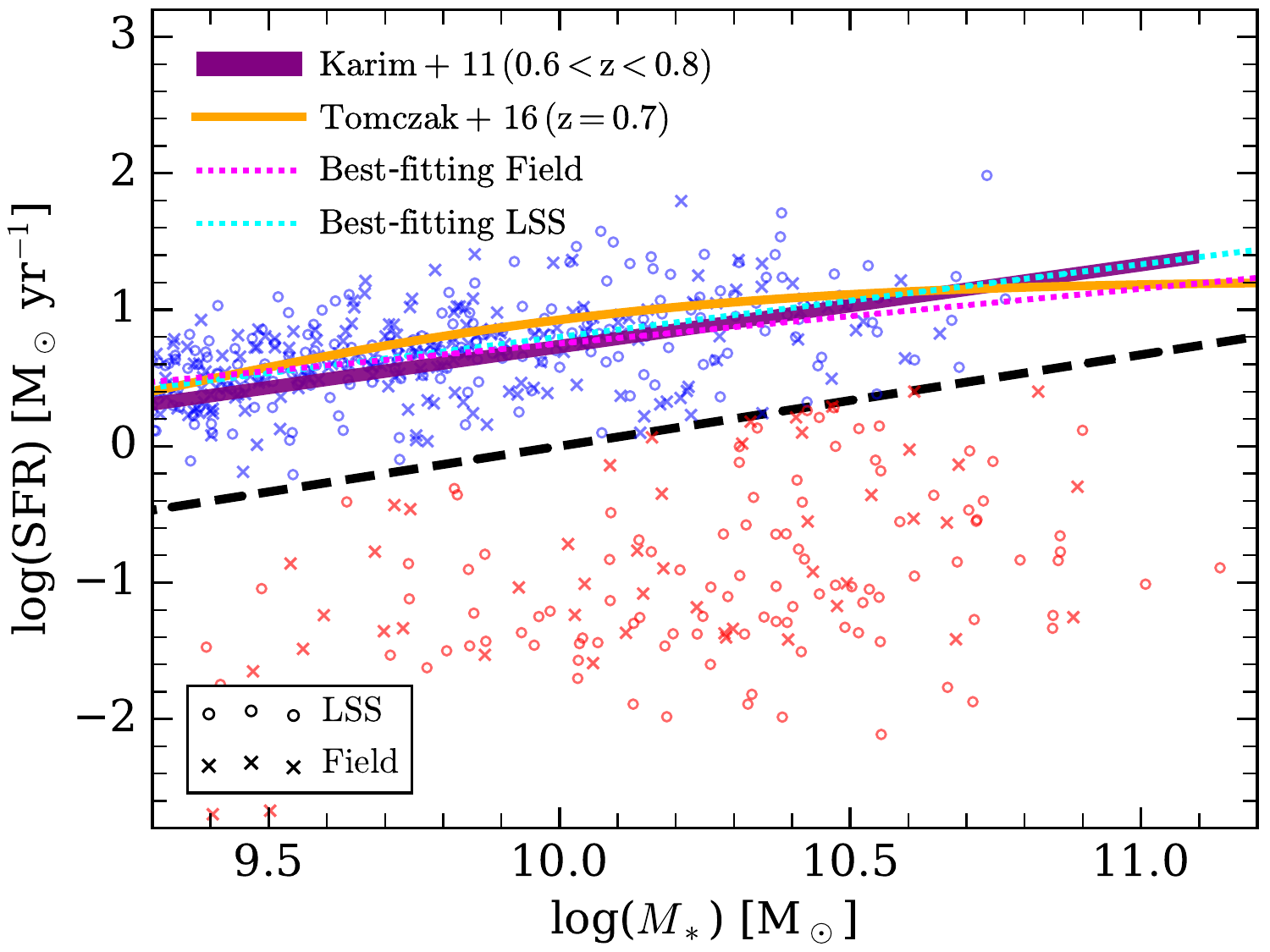}
         \caption{The $M_\ast$ versus SFR diagrams for galaxies at $0.69 \le  z \le 0.78$ in the ECDF-S field. The circles and crosses represent the galaxies in the LSS ($0.72 \le z \le 0.75$) and in the field ($0.69 \le z<0.72$ \&  $0.75<z \le 0.78$) respectively. The orange solid line is the SFMS relation at $z=0.7$ from \citet{Tomczak2016}. The purple solid line refers to the SFMS at $0.65 < z < 0.80$ from \citet{Karim2011}. The black dashed line is used to separate SFGs and QGs. The SFMS of the LSS and the field samples are plotted with the cyan and magenta dotted lines respectively.
         }

         \label{fig:06}
\end{figure}
The module \texttt{sfhdelayed} is used in our SED-fitting procedure with \texttt{CIGALE}.  The \texttt{BC03} stellar population model \citep{Bruzual2003} with a Chabrier initial mass function \citep[IMF;][]{Chabrier2003} is adopted in our fitting. The \texttt{dustatt\_modified\_starburst} dust attenuation curve \citep{Calzetti2000} and \texttt{Dale2014} dust emission \citep{Dale2014} are adopted. An E(B-V) factor of 0.44 is used to correct the different dust attenuation between old and young stellar populations \citep{Calzetti2000,Charlot2000,Wild2011,Qin2019}. For each model, a number of parameters are set to fit the observed data, which are listed in Table \ref{tab:sed}. Some examples of our SED-fitting results are shown in Figure~\ref{fig:05}.

Through SED-fitting, we obtain the stellar mass, SFRs and other physical parameters for 1\,402 galaxies at $0.69\le z \le 0.78$ in the ECDF-S. Some faint sources have larger reduced chi square ($\chi_r^2$) due to the observation errors, and these sources will be removed from our SED samples. The 1\,116 galaxies with $\chi_r^2<5$ are selected. Figure~\ref{fig:06} shows the $M_\ast-$SFR relation for the galaxies with $\chi_r^2<5$. It can be found that the star formation main sequence (SFMS) of our LSS and field samples are well consistent with those at $z=0.7$ reported in the literature, supporting that our stellar mass and SFR estimates from the SED-fitting are robust. Besides, the star forming galaxies (SFGs) in our LSS and field samples obey the similar SFMS. The two samples of SFGs do not show noticeable difference in SFR.

\subsection{Galaxy local environment and galaxy properties} \label{sec:Surface density and galaxies properties}

Many of the galaxy properties are found to be tightly correlated with environment. We examine the properties of galaxies as functions of environment in the LSS and field samples. For a galaxy, we employ the surface density $\Sigma_{10}$ as its local environment indicator. By definition, $\Sigma_{10}=11/(\pi r_{10}^{2}$), where $r_{10}$ is the projected distance from this galaxy to its 10th nearest neighbor. This local environment indicator has been used in many works \citep[e.g.][]{Dressler1980, Koyama2013, Pan2013, Yan2015}. For the 3 redshift slices we study in this work, we compute the galaxy local density ($\Sigma_{10}$) using all $R<24$\,mag galaxies (phot-$z$+spec-$z$) in that redshift slice. For our sample, $\Sigma_{10}$ spans a range from $\sim0.2$\,Mpc$^{-2}$ to $\sim 100$\,Mpc$^{-2}$.

\begin{figure}
	\centering
	\includegraphics[width=0.98\columnwidth]{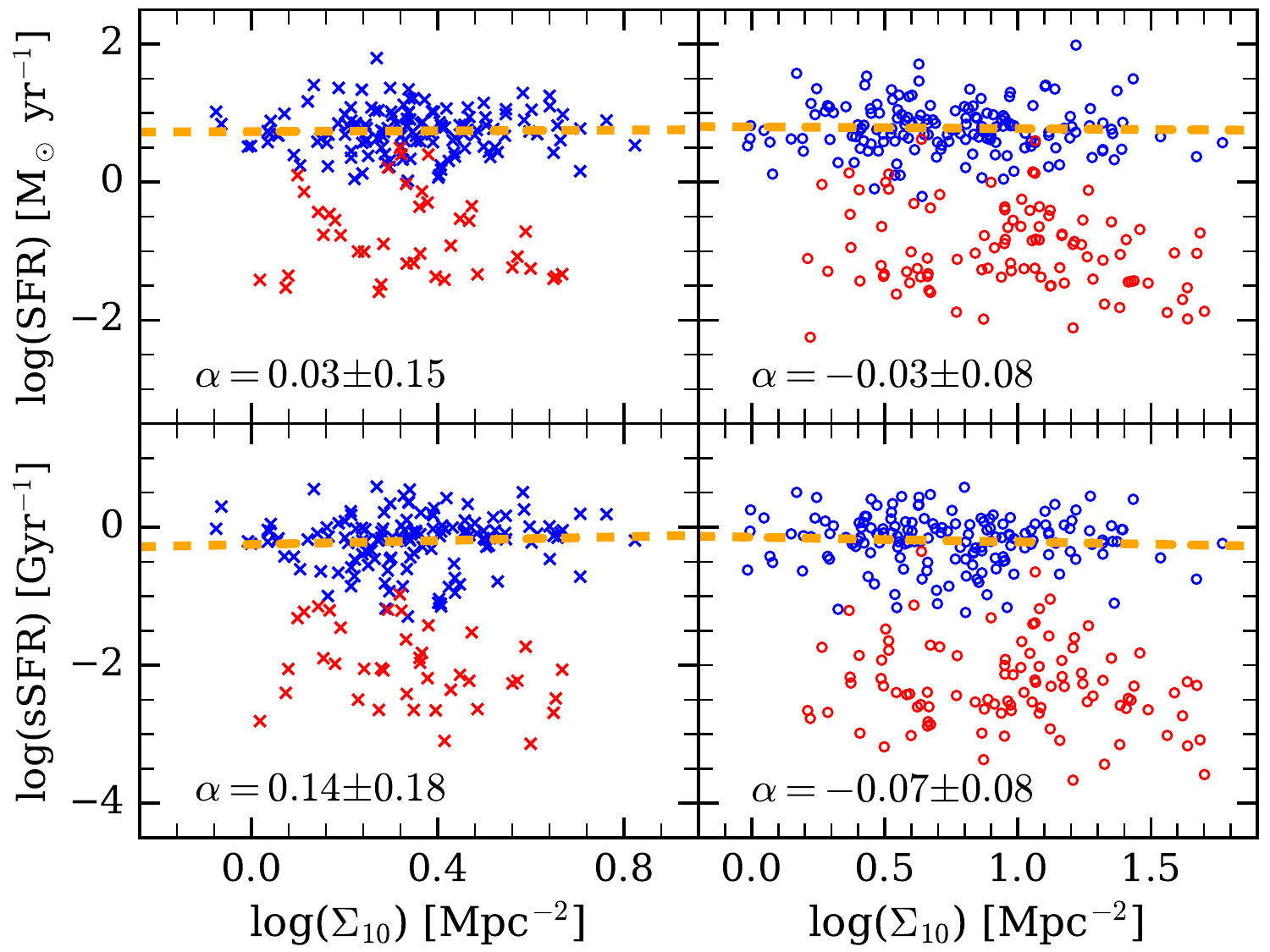}
         \caption{Comparison of galaxies between the field at $0.69\le  z < 0.72$ and $0.75 < z \le 0.78$ (left), and the LSS at $z = 0.735$ (right), in the diagrams of $\Sigma_{10}$ versus SFR (top), and $\Sigma_{10}$ versus sSFR (bottom). Blue symbols represent SFGs, and red symbols represent QGs. The golden dashed lines give the best-fitting relations to SFGs.
         }
         \label{fig:07}
\end{figure}

\subsubsection{Star formation properties in different environments}
\label{sec:Star formation properties in different environment}

Galaxy properties are usually correlated with stellar mass. Thus, a mass limited sample is demanded for many studies. In this section, we select $M_\ast>10^{9.3}$\,M$_\odot$ galaxies to study their star formation properties. This mass cut ensures that SFGs are mass complete within the probed redshift range \citep{Borch2006}.  SFGs and QGs form two peaks in the mass$-$SFR$-$density plot \citep[e.g.][]{Renzini2015}. We use a line along the lowest galaxy number density between these two peaks to separate these two populations. As such, galaxies with $\log(\rm {SFR})>0.65\log(M_\ast/\rm M_\odot)-6.7$ are classified as SFGs, as shown in Figure~\ref{fig:06}.  Figure~\ref{fig:07} shows the relationships between the density $\Sigma_{10}$ and star formation activity (described by SFR and sSFR $\equiv$ SFR/$M_\ast$) for the field and LSS samples. The blue crosses (circles) and red crosses (circles) represent SFGs and QGs in the field (LSS), respectively. The orange dashed lines draw the best-fitting relations for the SFGs. Note that $\Sigma_{10}$ spans about two orders of magnitude in the LSS, while it only ranges up to 10\,Mpc$^{-2}$ in the field. The high end of $\Sigma_{10}$ in the LSS is comparable to that in massive clusters of galaxies in the local Universe \citep[e.g.][]{Yan2015}.

 SFR of SFGs does not change with increasing $\Sigma_{10}$ both in the LSS and the field, giving a slope of $\alpha= 0.03\pm0.15$ and $\alpha= -0.03\pm0.08$, respectively.  \citet{Vulcani2010} finds that denser regions tend to contain higher fraction of massive galaxies. To remove the stellar mass dependence, we investigate sSFR against $\Sigma_{10}$ and show the results in the bottom panels of Figure~\ref{fig:07}. The best-fitting relation obeys $\log(\rm{sSFR}) = 0.14 \pm 0.18  \times  \log(\Sigma_{10}) + 0.25 \pm 0.07$ and $\log(\rm{sSFR}) = -0.07 \pm 0.08 \times  \log(\Sigma_{10}) + 0.14 \pm 0.07$ for SFGs in the field and LSS samples, respectively. One can see that sSFR only slightly increases with increasing $\Sigma_{10}$ for SFGs in the field sample, and sSFR is nearly independent of $\Sigma_{10}$ for SFGs in the LSS.

Within the fitting uncertainties, the sSFR$-\Sigma_{10}$ relations are nearly identical for both of the two samples.  Differing from the star formation$-$density relation seen among local galaxy clusters. Our results reveal that SFGs at $z\sim0.7$ in the ECDF-S field demonstrate no star formation dependence on the local environment traced by $\Sigma_{10}$, indicating that the local environment does not play a significant role in either enhancing or suppressing star formation in SFGs, at least for our LSS and field samples.
Compared with the general field, the $z=0.735$ LSS provides a cosmic environment that exhibits little or no effect on the suppression of star formation in SFGs. A similar result is also reported by \citet{Poggianti2008} at the same redshift.


\begin{figure}
	\centering
	{\includegraphics[width=0.98\columnwidth]{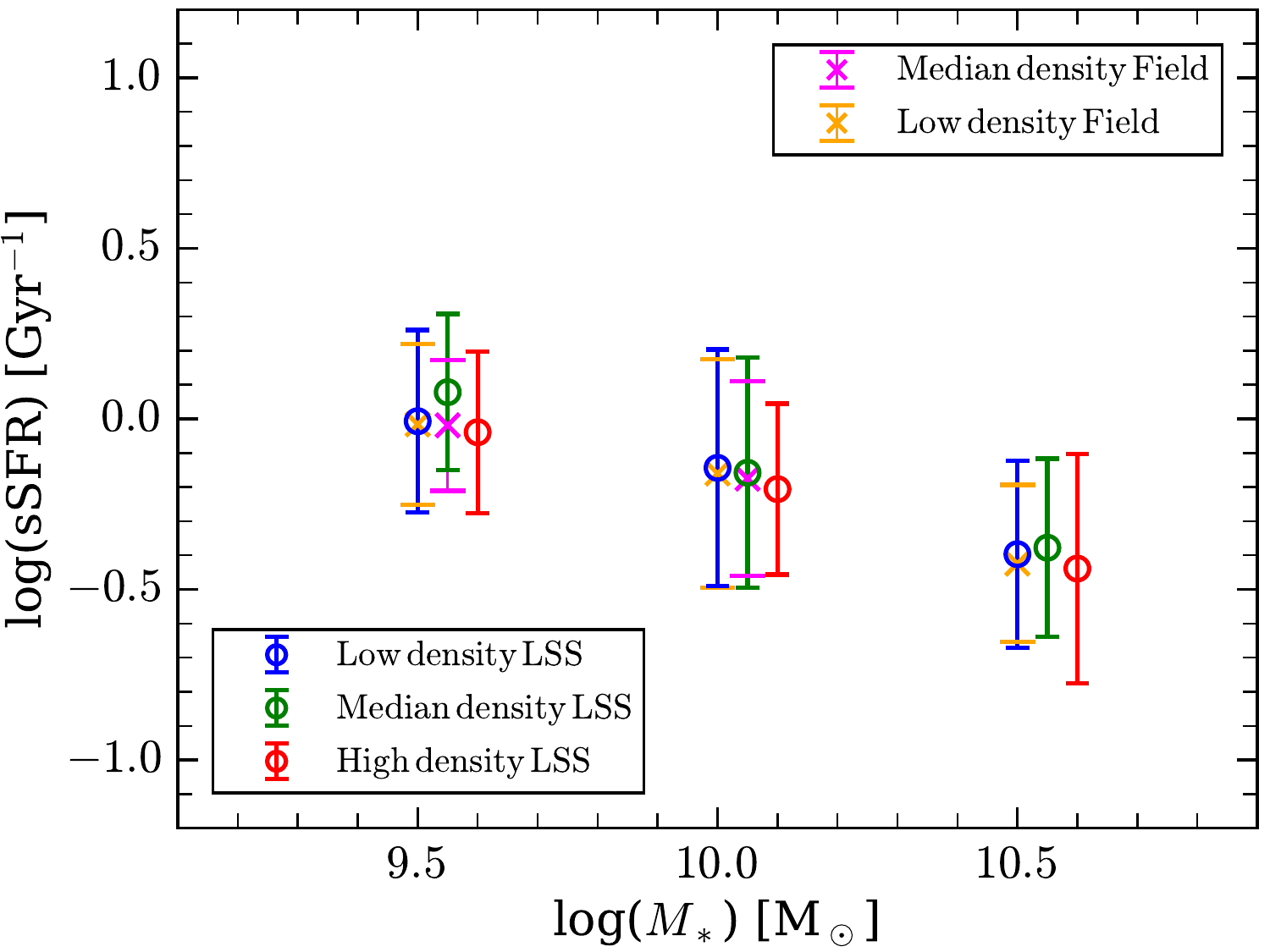}}
         \caption{The median sSFR as a function of stellar mass for SFGs split into stellar mass and local density bins in the $z = 0.735$ LSS. The three local-density bins refer to low- (blue circles), medium- (green circles) and high-density (red circles) in the LSS. The magenta and orange crosses refer to the median- and low-density in the field. The error bar of every data point marks the sSFR range including 68\,per\,cent of galaxies in that bin.
         }
         \label{fig:08}
\end{figure}


As already known, sSFR of SFGs also depends on stellar mass. To separate the dependence of sSFR on stellar mass and $\Sigma_{10}$, we further divide the LSS sample into subsamples according to their $\Sigma_{10}$ and stellar mass, and investigate their distributions in the sSFR$-M_\ast$ diagram. We divide the LSS sample into low-density ($\log \Sigma_{10}<0.48$), medium-density ($0.48<\log \Sigma_{10}<0.87$) and high-density ($\log \Sigma_{10}>0.87$) subsamples, ensuring that each subsample contains similar number of galaxies. In Figure~\ref{fig:08},  the median sSFR are plotted against the median of stellar mass for three subsamples. Note that our field samples contain few galaxies in the high-density environment. It is obvious that sSFR decreases with stellar mass but the local density ($\Sigma_{10}$) in the first place.
The three subsamples follow the same $M_\ast-$sSFR relation. Therefore, we conclude from Figure~\ref{fig:07} and Figure~\ref{fig:08} that galaxy local environment ($\Sigma_{10}$) does not induce significant influence on star formation of SFGs in the $z=0.735$ LSS.

\begin{figure}
	\centering
	{\includegraphics[width=\columnwidth]{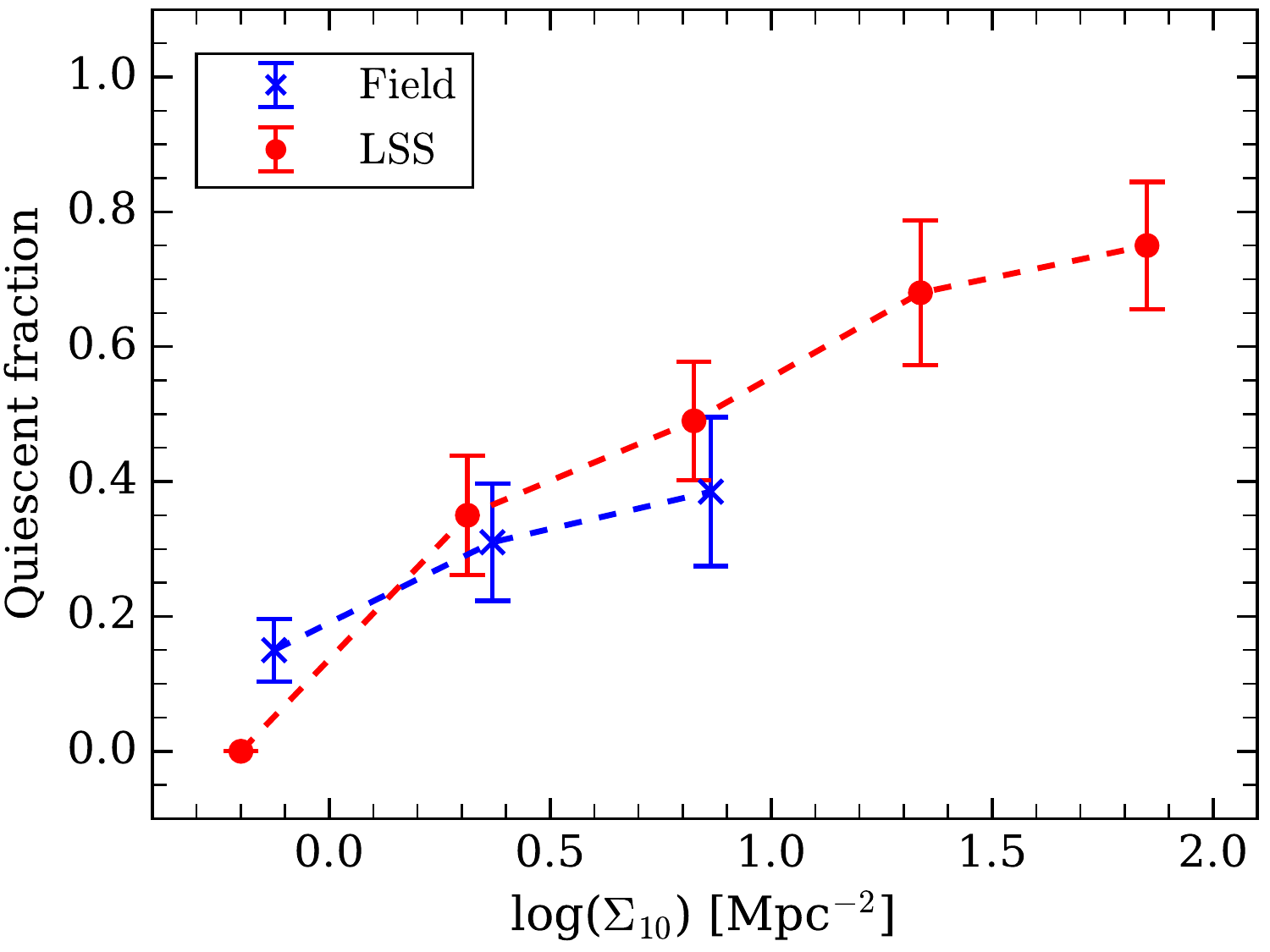}}
         \caption{Fraction of QGs as a function of  local density for the galaxies with $M_\ast>10^{10}$\,M$_{\odot}$ in the field (blue crosses) and the $z = 0.735$ LSS (red dots) samples. The error bars indicates shot noises.}
         \label{fig:09}
\end{figure}

The environments have influence not only on the star formation properties, but also on the transformation of SFGs into QGs. Therefore, the study of the QG fraction ($f_{\rm QG}$)  in the field and in the LSS can better understand how the environment quench the star formation. The selection criterion of $R<24$\,mag at $\sim0.7$ is incomplete for QGs with $M_\ast<10^{10.3}$\,M$_\odot$ \citep{Borch2006}. Considering the lack of massive galaxies in our sample, we counted the  $f_{\rm QG}$ using the galaxies with $M_\ast>10^{10}$\,M$_\odot$.

Figure \ref{fig:09} shows the $f_{\rm QG}$ as a function of local density. In the LSS, the $f_{\rm QG}$ can reach $\sim 70$\,per\,cent in the high density region. This illustrates that in the most dense regions, the quenching of star formation is linked to environmental effects.
The $f_{\rm QG}$ of the field sample is similar to that in the low-density region of the LSS and increase with the increase of local density both in the field and in LSS.  It means that the LSS has little effect on quenching galaxies in the low-density region around it. This is like putting several dense regions in a general field with no effect on the surrounding galaxies. In other words, the $f_{\rm QG}$ is influenced by local environment rather than the large-scale environment, at least in our work.

\subsubsection{Stellar mass and S\'ersic index} \label{sec:Stellar mass in different environment}

\begin{figure}
	\centering
	\includegraphics[width=\columnwidth]{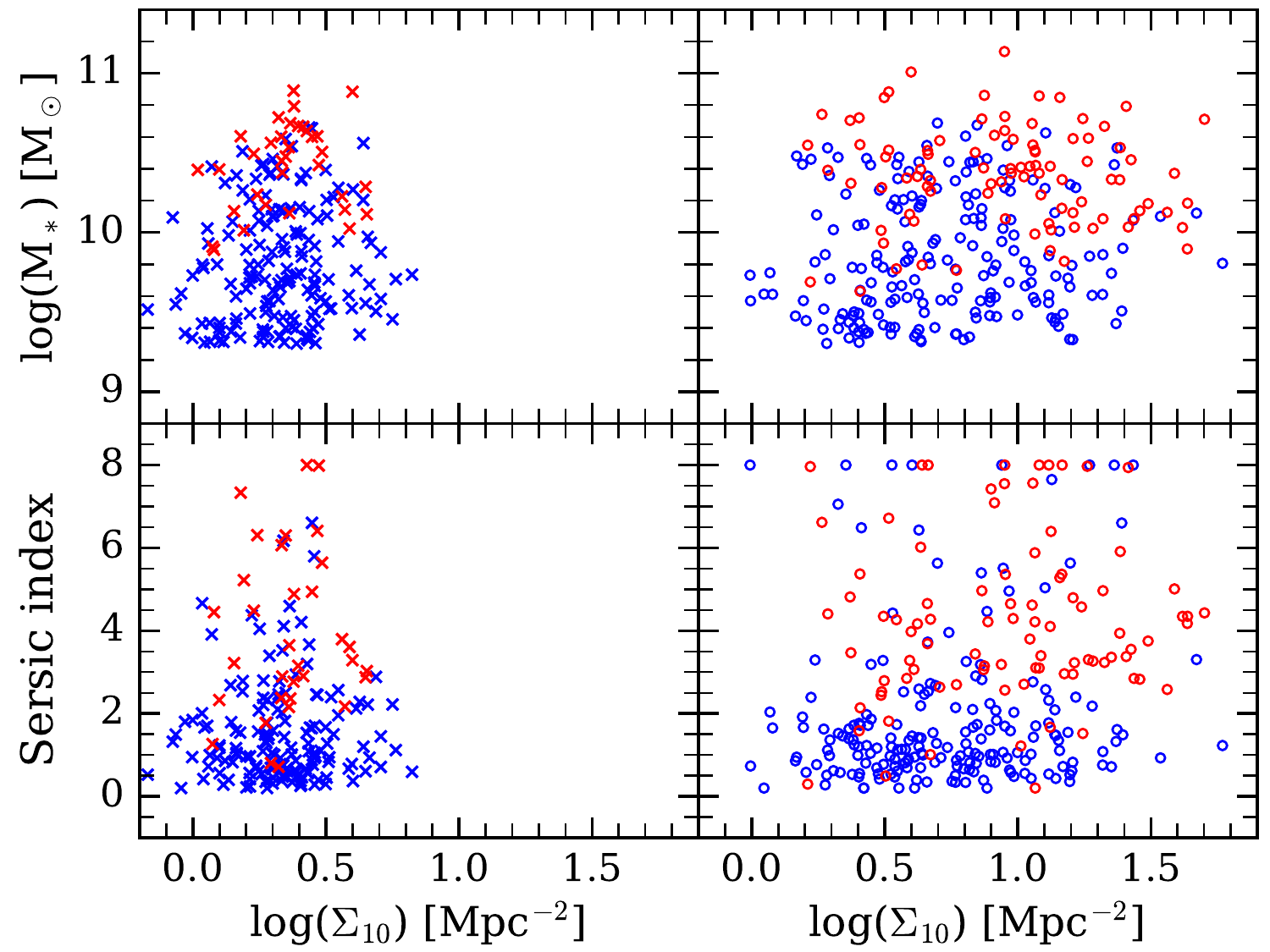}
         \caption{Galaxy stellar mass (top) and S\'ersic index (bottom) as a function of local density for galaxies in the field (left) and in the LSS (right). The symbols are the same as in Figure~\ref{fig:07}.}
         \label{fig:10}
\end{figure}

We examine the relations between galaxy local density,  stellar mass and structural parameter for the galaxies with $M_\ast>10^{9.3}$\,M$_\odot$ .  Here S\'{e}rsic index ($n$) is used as the structural parameter to characterize galaxy morphology and light concentration. The GEMS catalog provides structural parameter measurements  for nearly 10\,000 galaxies \citep{Rix2004}. We cross-match our EMRC catalog with the GEMS catalog and obtain S\'{e}rsic index for 72\,per scent of the total sample galaxies. Figure~\ref{fig:10} shows  the distributions of stellar mass and S\'{e}rsic index in relations to $\Sigma_{10}$ for our sample galaxies in the field and the $z=0.735$ LSS.  As can be seen,  in the field no clear tendency is found between galaxy stellar mass and local density. Massive galaxies with $M_\ast>10^{10}$\,M$_\odot$ in the field  reside in the low- and medium-density environments, while in the LSS, galaxies of different stellar masses appear in all three types  (low- medium- and high-density) of local environments. Only very high-density environment ($\log\Sigma_{10}>1.5$) in the LSS preferentially hosts massive galaxies with $M_\ast>10^{10}$\,M$_\odot$.  From the bottom panels of Figure~\ref{fig:10},  no correlation is found between $n$ and $\Sigma_{10}$ in either the field or the LSS.  We emphasize that in the LSS galaxies with $n>3$ can be found in all three types of local environment, and they do not favour the medium- and high-density environment.

\subsubsection{The morphology$-$density relation in cluster-B} \label{sec:The density-morphology relation in cluster B}

We examine whether the morphology$-$density relation among local galaxy clusters is already in place at intermediate $z$. To do so, a large sample of member galaxies with spec-$z$ and high-resolution imaging are needed. In the $z=0.735$ LSS, only Cluster-B hosts member galaxies with a high completeness in spec-$z$ (93\,per\,cent within $r/R_{200}<2$), and \textit{HST}/ACS multi-band imaging data from the GOODS-S survey are available.

In order to study the morphology$-$density relation in the inner ($r/R_{200}<2$) and the outer($2<r/R_{200}<3$) region of cluster-B,
we selected a sample of 101 galaxies from cluster-B within $r/R_{200}<3$ that with \textit{HST}/ACS F850LP imaging data.
These galaxies are then visually classified into four morphological classes: disc, elliptical, merger and peculiar. In our classification, discs include spiral, lenticular and edge on galaxies. Galaxies that exhibit tidal tails, double nuclei, or significant merger remnants are categorized into mergers. Galaxy pairs are not included in our merger sample, because the high number density of galaxies in clusters brings in very severe  projection effect, resulting in a high misidentification rate of galaxy pairs. In addition, the high relative velocity between galaxies in clusters makes it difficult for galaxy pair to merge.  The peculiar class includes irregular and morphologically disturbed galaxies. In total, there are 39 elliptical (39\,per\,cent), 26 disc (26\,per\,cent), 13 merger (13\,per\,cent) and 23 peculiar (23\,per\,cent) galaxies in our sample.

\begin{figure}
	\centering
	\includegraphics[width=\columnwidth]{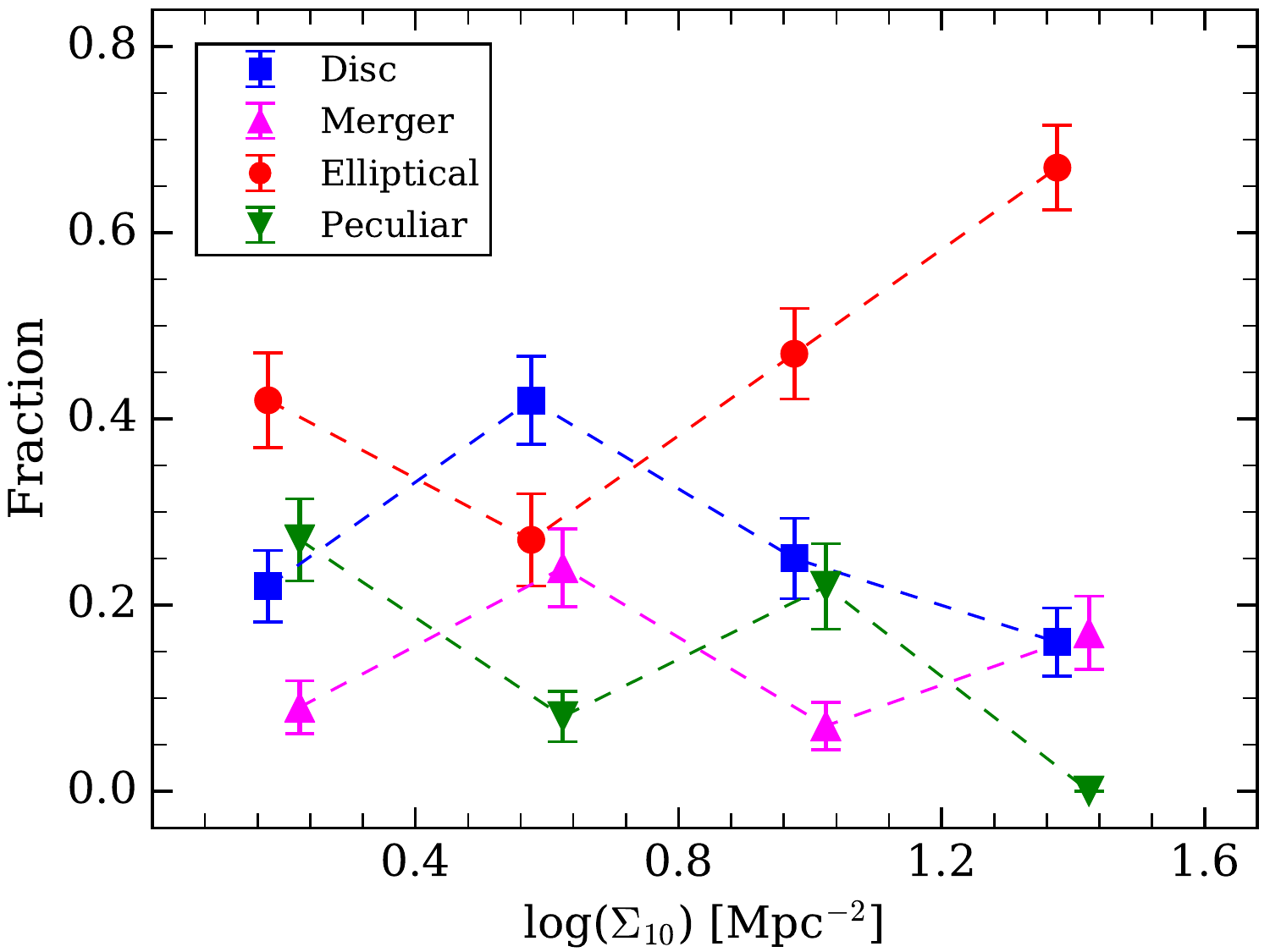}
         \caption{The galaxy fraction of different morphology types in different environments. The error bars indicates shot noises.}
         \label{fig:11}
\end{figure}

We divide the environment indicator $\Sigma_{10}$ into four bins and then count galaxies of given morphological types in each $\Sigma_{10}$ bin. The results are shown in Figure~\ref{fig:11}.  It is obvious that the fraction of elliptical is higher toward denser environment, and discs show an opposite trend. These results are similar to those found in local clusters of galaxies. Interestingly, the fraction of disc and mergers both peak at the density range of $0.4<\log(\Sigma_{10})<0.8$. This may partially due to the applied morphological classification. In our classification,  we choose the mergers based on the tidal features, which is generated by the merger of disc galaxies. Therefore, the fraction of mergers is relatively high in the environment with high-fraction disc galaxies. For peculiar galaxies, they peak at $0<\log(\Sigma_{10})<0.4$ and $0.8<\log(\Sigma_{10})<1.2$. The low-density peak is expected, because irregular galaxies are mostly low-mass galaxies and tend to reside in low-density environments (outer). In the high-density environment (inner) , strong interactions between galaxies can cause strong disturbance, also producing a significant fraction of peculiar galaxies.

\section{Discussion} \label{sec:Discussion}

The $z=0.735$ LSS in the ECDF-S filed spanning $23.9\times 22.7$\,cMpc$^2$ over the observed area is characterized by a low velocity dispersion of $494\pm24$\,km\,s$^{-1}$ along the line of sight, likely being a thin sheet consisting of five galaxy clusters and surrounding filaments. This LSS contains galaxies at a density $\gtrsim 3.9$ times that of the general fields, suggestive of an overdense cosmic environment that is comparable to galaxy clusters at the same epoch \citep[e.g.][]{Chiang2013}. Our results reveal that SFGs in this dense structure are indistinguishable from those in the field at the same redshift in terms of their SFR, as shown in Figure~\ref{fig:07}. This finding differs from the expectation that galaxy evolution in dense environments tends to be accelerated \citep[e.g.][]{Thomas2005}, and cluster SFGs exhibit a lower sSFR relative to these in the field at intermediate redshifts \citep{Erfanianfar2016, Jian2018}.  However, no environmental dependence of galaxy star formation was also reported \citep[e.g.][]{Randriamampandry2020}. We notice that previous studies on environmental effects were mostly concentrated on cluster-scale environment, i.e., the local environment mentioned in our work. The cosmic environment beyond the cluster scale has seldom been addressed. We argue that the controversy on environmental effects on galaxy star formation is likely attributed to the omission of the cosmic environment.

Our striking finding that no environmental dependence is seen for the SFMS as well as the sSFR$-$density relation in the $z=0.735$ LSS suggests that the cosmic environment overtakes the local environment in remaining galaxy star formation, at least in some cosmic large-scale structures. We point out that the $z=0.735$ LSS enables to effectively supply gas into member galaxies in various local environments and sustains  star formation at the level of the field.

\subsection{The robustness of the density$-$sSFR relation } \label{sec:Density indicator}
We demonstrate that adoption of different density indicators will not change our results relied on galaxy local density given in Figure~\ref{fig:07} and Figure~\ref{fig:10}. \citet{Muzzin2012} used the cluster centric distance as a proxy of density. The smaller the cluster centric distance, the higher the number density of galaxies. This density indicator is suitable for clusters or groups. Some works use the relative density, setting the lowest density region to zero and the highest density region to unity to facilitate the comparison of different fields \citep{Scoville2013}. The most commonly used method for galaxy density is the two- or three-dimensional number density of galaxies within the radius of the $i$-th nearest galaxy \citep{Elbaz2007, Patel2011, Koyama2013}.  We use $\Sigma_{10}$ as the indicator for galaxy local density. Our choice of $i=10$ is optimized for a better comparison of the sSFR$-$density relation between the field and the LSS with given sample galaxies.  It is a balance between the density resolution and the sensitivity to small-scale substructures.

\begin{figure}
	\centering
	\includegraphics[width=\columnwidth]{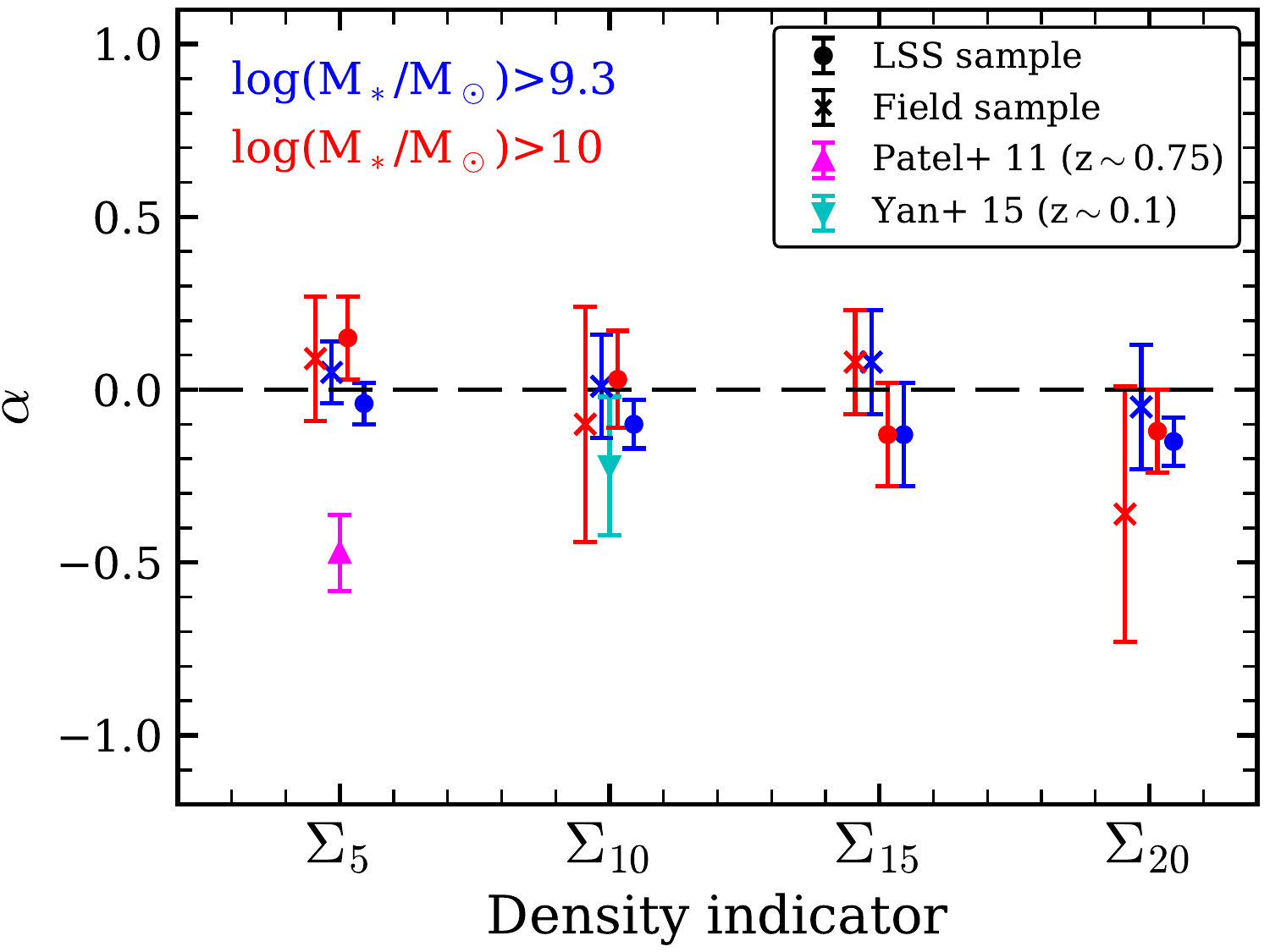}
         \caption{ The slope of the density$-$sSFR relation versus density indicator. The blue and red symbols represent the samples with  $M_\ast>10^{9.3}$\,M$_\odot$ and  $M_\ast>10^{10}$\,M$_\odot$  respectively. The crosses are the field samples and the dots are the LSS samples. The magenta and cyan symbols represent the results from \citet{Patel2011} and \citet{Yan2015}.
         }
         \label{fig:12}
\end{figure}

We test the density indicators with $i=5, 15,20$ and examine the influence on the sSFR$-$density relation. Figure~\ref{fig:12} illustrates the slope ($\alpha$) of the sSFR$-$density relation obtained with the four density indicators. The results from the literature are also plotted. The blue and red points show the galaxies with $M_\ast>10^{9.3}$\,M$_\odot$ and $M_\ast>10^{10}$\,M$_\odot$, respectively. The crosses and dots represent the filed and LSS samples. Triangles show the results collected from literatures \citep{Patel2011,Yan2015}. The error bars indicate the best-fitting errors of the sSFR$-$density relations.  It can be seen that SFGs in the LSS and the field share a similar $\alpha$ at given density indicators.  sSFR$-$density relations are not affected by different mass limited samples. When  using $\Sigma_{15}$ or $\Sigma_{20}$ as the density indicator, the density range is significantly compressed and the scaling relations suffer from large uncertainties.

We point out that our SFR measurements do not induce systematics that could bias our results on either the SFMS or the sSFR$-$density relation.  We have shown that our SFMS is well consistent with those at the same redshift given in the literature, verifying that our SFR measurements are robust.  On the one hand, as \citet{Patel2011} pointed out,  the systematics induced by different SFR indicators only affect the normalization of the SFMS and the sSFR$-$density relation for both the field and the LSS. On the other hand, SED-based SFRs have a lower detection limit than that of the [O\,{\footnotesize{II}}]-based SFRs. \cite{Vulcani2010} demonstrated a detection limit of  0.3\,M$_\odot$ yr$^{-1}$ for the [O\,{\footnotesize{II}}]-based SFRs.  The SFRs of our SFGs sample are larger than 0.5\,M$_\odot$ yr$^{-1}$, which have not reached the detection limit. Thus our conclusions based on the comparison between the two samples are not affected.

\subsection{The effects of environment} \label{sec:Galaxy evolution in the LSS}

Investigating the star formation properties of galaxies in the LSS is crucial for understanding the roles of environment on galaxy evolution. As shown in Figure~\ref{fig:07}, sSFR is only weakly correlated with local density ($\Sigma_{10}$)  for SFGs in both the LSS and the field. There have been many works studying the correlations between star formation properties and environment from low-$z$ to high-$z$, but giving controversial conclusions.  \citet{Cooper2008} reported a negative sSFR$-$density correlation based on SDSS and DEEP2 data. However, some other studies reported a positive or no correlation at $z<1$ \citep[e.g.][]{McGee2011,Muzzin2012,Koyama2013}. \citet{Patel2011} found that sSFR decreases by 1\,dex from the lowest to the highest environment at $z\sim 0.75$. For a comparison, our results show a decrease of sSFR of only $\sim 0.2$\,dex, suggesting that high density environment in the LSS has little effect on galaxy sSFR. We suspect that the discrepancy of the sSFR$-$density relations from those studies may be caused by the failure to take global environment and local environment into account. Although the dense environment does have some influence on star formation activity, our work indicates that star formation activity in the LSS is similar to that in the field. This likely indicates that interactions play a similar role both in the LSS and in the field at the same local density.

In the $z=0.735$ LSS, we find that there are some filamentary substructures connecting five clusters. We suspect that the filamentary structures are transporting gas and galaxies into these clusters, which maintains the star formation activity of SFGs in the LSS. In addition, as indicated by the spec-$z$ distribution, the line-of-sight velocity dispersion of the LSS is only $\sigma \sim 500$\,km\,s$^{-1}$, which is far lower than that of the local massive structures such as the Coma cluster ($\sigma>1000$\,km\,s$^{-1}$). The global environment may affect the heating efficiency of gas in the structures through stellar and active galactic nucleus feedback or gravitational heating.  For a thin sheet, the feedback energy and hot gas from galaxies escape easily from the sheet and gas in the sheet is much less affected.  Moreover, most of the substructures in the $z=0.735$ LSS seem have not reached dynamically relaxed and thus gravitational heating to the IGM is negligible. The gas cooling and accretion into galaxy halos is still as efficient as that in the general fields. On the other hand, if the global environment lead the feedback injection and/or gravitational heating to the IGM more efficient, the gas cooling into a galaxy halo would become less efficient and galaxy star formation would decline compared to the general field at the same epoch.

We note that there has been a high $f_{\rm QGs}$ in the denser environment at $z=0.7$. \citet{van der Burg2020} points out that the QG fraction is higher in clusters than that in fileds at $z\sim$1.25.  At high redshift, \citet{Wang2016} points out that the $f_{\rm QGs}$ is similar to that in the field by studying a protocluster at $z=2$.  This means that the $f_{\rm QGs}$ increases rapidly at $z<2$. We believe that the cluster environment is critical for the transition from SFGs to QGs since $z=2$.
As already known, external or internal-related mechanisms (known as `environment quenching' and `mass quenching') are both responsible for star formation quenching in galaxies \citep[e.g.][]{Peng2010, Old2020}. The detailed discussion of `mass quenching' is beyond the scope of this paper. Our results also indicate that the $f_{\rm QGs}$ in the low-density environment of the LSS is similar to that in the field, indicating that the quenching of star formation is not sensitive to the large-scale environment, at least for our sample. In the medium-density environment, galaxy interaction frequently occurs, which can effectively consume or expel the gas in galaxies and may be an important mechanism for quenching. In the core of clusters, the interactions between galaxies and intra-cluster medium (ICM) can effectively strip gas out of a galaxy and lead to quenching. Our results also suggest that galaxy clusters have little influence on the galaxies in the surrounding filaments. 

\subsection{The $z=0.735$ LSS: a structure under assembly} \label{sec:The formation and evolution of LSS}

In the $z=0.735$ LSS, we identify five clusters and study the properties for four of them. Although we use large samples of member galaxies with spec-$z$ to compute  $R_{200}$ and $M_{200}$,  and obtain a similar result as \cite{Dehghan2014}, we remind that these results should be seen as the lower limits since the calculation of $R_{200}$ and $M_{200}$ is made under the assumption that these clusters are virialized systems. From our visual inspection, Cluster-C and Cluster-D contain substructures, which are considered to be independent groups in \citet{Dehghan2014}. Cluster-A is ellipsoidal in morphology and has a very dense core, so it is the most likely cluster to have reached dynamic equilibrium. Cluster-B locates near the centre of the LSS and connects with other clusters. It hosts the largest number of member galaxies but smallest $R_{200}$ and $M_{200}$.

In Cluster-B, 40\,per\,cent of the member galaxies are elliptical, and the elliptical fraction increases toward denser regions. In this cluster, the fraction of merging galaxies is up to 13\,per\,cent. The actual fraction could be even higher since the selection of merging galaxies based on morphology would miss the merging between elliptical galaxies. \citet{Oh2018} studied low-$z$ clusters and found that the merger fraction reaches 20\,per\,cent. We argue that it is not surprising that a high fraction of merging galaxies is seen in unrelaxed clusters, since the galaxy interaction rate is expected to be high during the collapse of substructures of clusters.

\section{Summary} \label{sec:Summary}

Using the publicly available phot-$z$ and spec-$z$ data combined with our new spectral observations with Magellan/IMACS MOS, we re-examine a large-scale structure at $z=0.735$ in the ECDF-S field. We select 732 galaxies with $R<24$\,mag at $0.72 \le z \le 0.75$ belonged to the  $z=0.735$  structure.  We carry out broad-band SED-fitting with \texttt{CIGALE} for galaxies at $0.69 \le z \le 0.78$ to derive their physical parameters, including stellar mass and SFR. We pick up galaxies at $0.69 \le z<0.72$ and $0.75<z \le 0.78$ as the field sample for a comparison with the LSS sample. The substructures of the LSS, sSFR$-$density relation, $M_\ast-$density relation,  and morphology$-$density relation are addressed.  Our results are summarized as follows:
\begin{itemize}
\item[\textbf{1.}] The $z=0.735$ LSS  spanning $23.9\times 22.7$\,cMpc$^2$ over the observed area consists of 732 member galaxies with $R<24$\,mag and a line-of-sight velocity dispersion of $494\pm24$\,km\,s$^{-1}$,  yielding a galaxy number density $\gtrsim 3.9$ times that of the field,  and suggestive of an overdense cosmic environment that is comparable to galaxy clusters at the same epoch.

\item[\textbf{2.}] Five substructures are identified in the LSS based on the spatial distribution of member galaxies. Considering their physical scale, these substructures are classified as galaxy clusters. Two of the clusters contain substructures (50\,per\,cent),  which are comparable to low-$z$ clusters.

\item[\textbf{3.}] SFGs in both the $z=0.735$ LSS and the field obey the similar sSFR$-$density relation, showing a weak dependence of sSFR on the local density. We demonstrate that this weak dependence is controlled by galaxy stellar mass-related scaling relation. Star formation activity of SFGs in either the field or the LSS exhibits no dependence on the local environment traced by $\Sigma_{10}$.  However, the fraction of QGs increases with the local density both in the LSS and in the field. The $f_{\rm QG}$ in the low-density region of the LSS is similar to that of the galaxies in the field, which means that the LSS does not influence the quenching of the low-density-environment galaxies around it.

\item[\textbf{4.}] The morphology$-$density relation is presented for one cluster based on \textit{HST}/ACS imaging. The fraction of merging galaxies is found to be relatively high, which is interpreted as a natural consequence during the dynamical relaxation process of a galaxy cluster.
\end{itemize}
Our results suggest that the $z=0.735$ LSS is still in an assembly phase and far from being virialized. The SFGs in the dense regions are likely transported from the surrounding filamentary substructures.  We thus stress that the cosmic environment overtakes the local environment in remaining galaxy star formation in at least some large-scale structures like the $z=0.735$ one in the ECDF-S field.

\section*{Acknowledgements}
The authors thank the anonymous referee for the useful and detailed comments and suggestions. This work is supported by the National Key R\&D Program of China (2017YFA0402703), the National Science Foundation of China (11773076 and 12073078),  the Major Science and Technology Project of Qinghai Province (2019-ZJ-A10), the science research grants from the China Manned Space Project with NO. CMS-CSST-2021-A02 and CMS-CSST-2021-A07, and the Chinese Academy of Sciences (CAS) through a China-Chile Joint Research Fund (CCJRF \#1809) administered by the CAS South America Centre for Astronomy (CASSACA). This research made use of \texttt{Astropy}, a community-developed core Python package for Astronomy \citep{AstropyCollaboration2013}.
\section*{Data Availability}
The data underlying this article will be shared on reasonable request to the corresponding author.




\appendix

\section{Some extra material}
\label{sec:appendix}
Figure \ref{fig:13}: Example spectra in our observations.  \\
Table \ref{tab:observation}:  $Column$\,1: slits ID in our observations; $Column$\,2-6: objects id, coordinates, R-band magnitudes and photometry redshifts come from COMBO-17 catalog; $Column$\,7: spectroscopic redshifts.


{
\begin{figure*}
	\centering
	\includegraphics[width=0.3\textwidth]{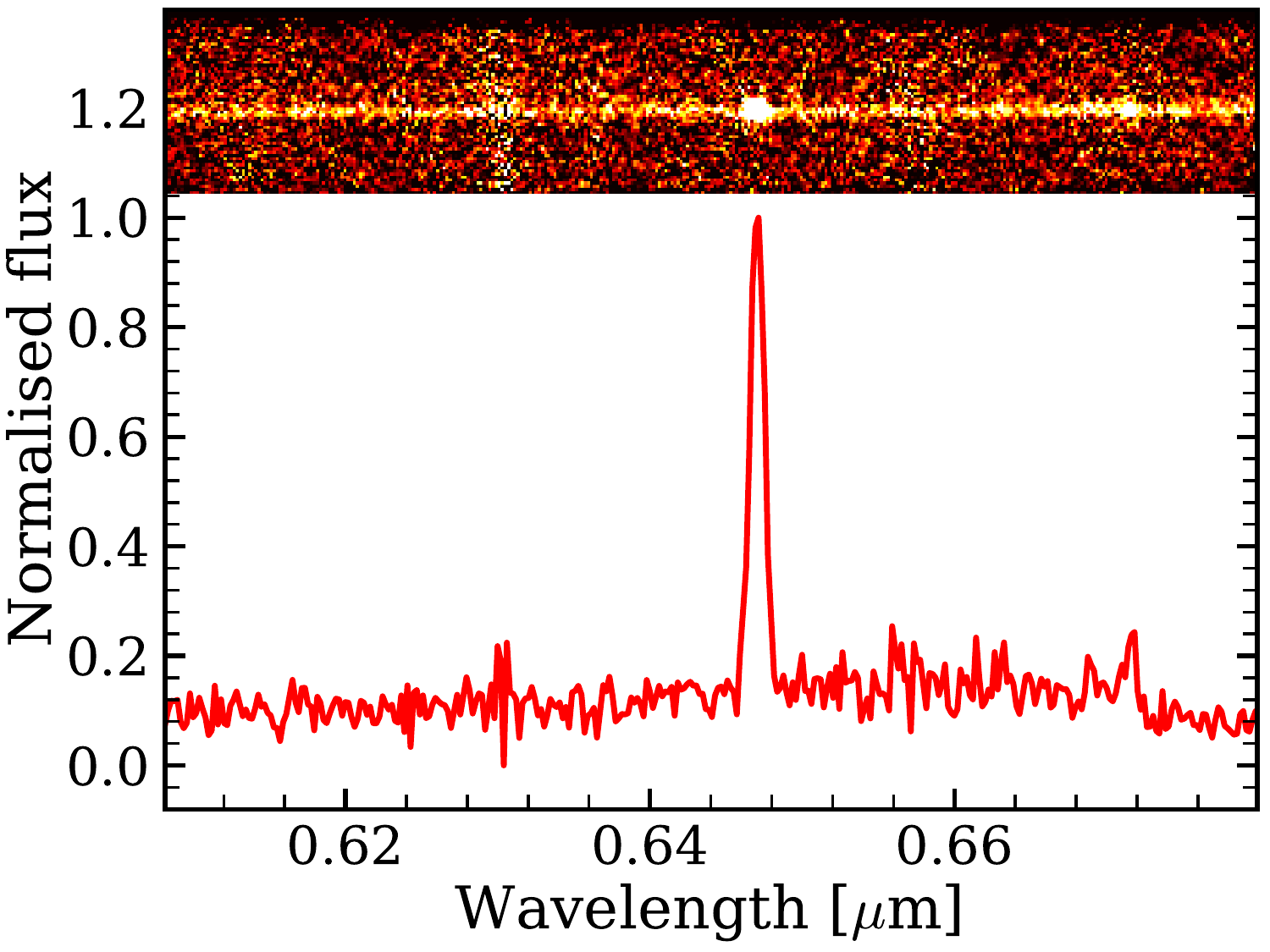}
	\includegraphics[width=0.3\textwidth]{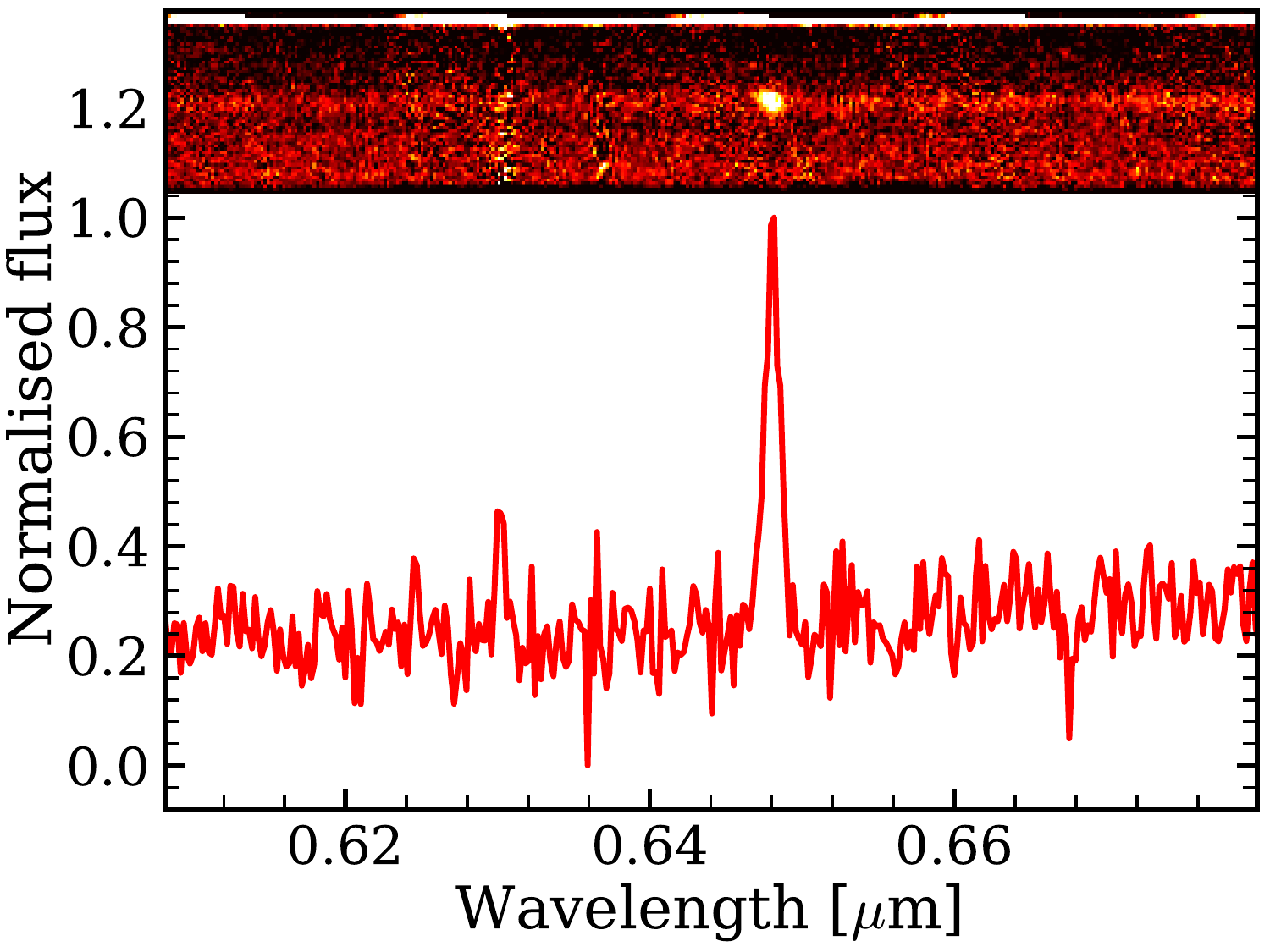}
	\includegraphics[width=0.3\textwidth]{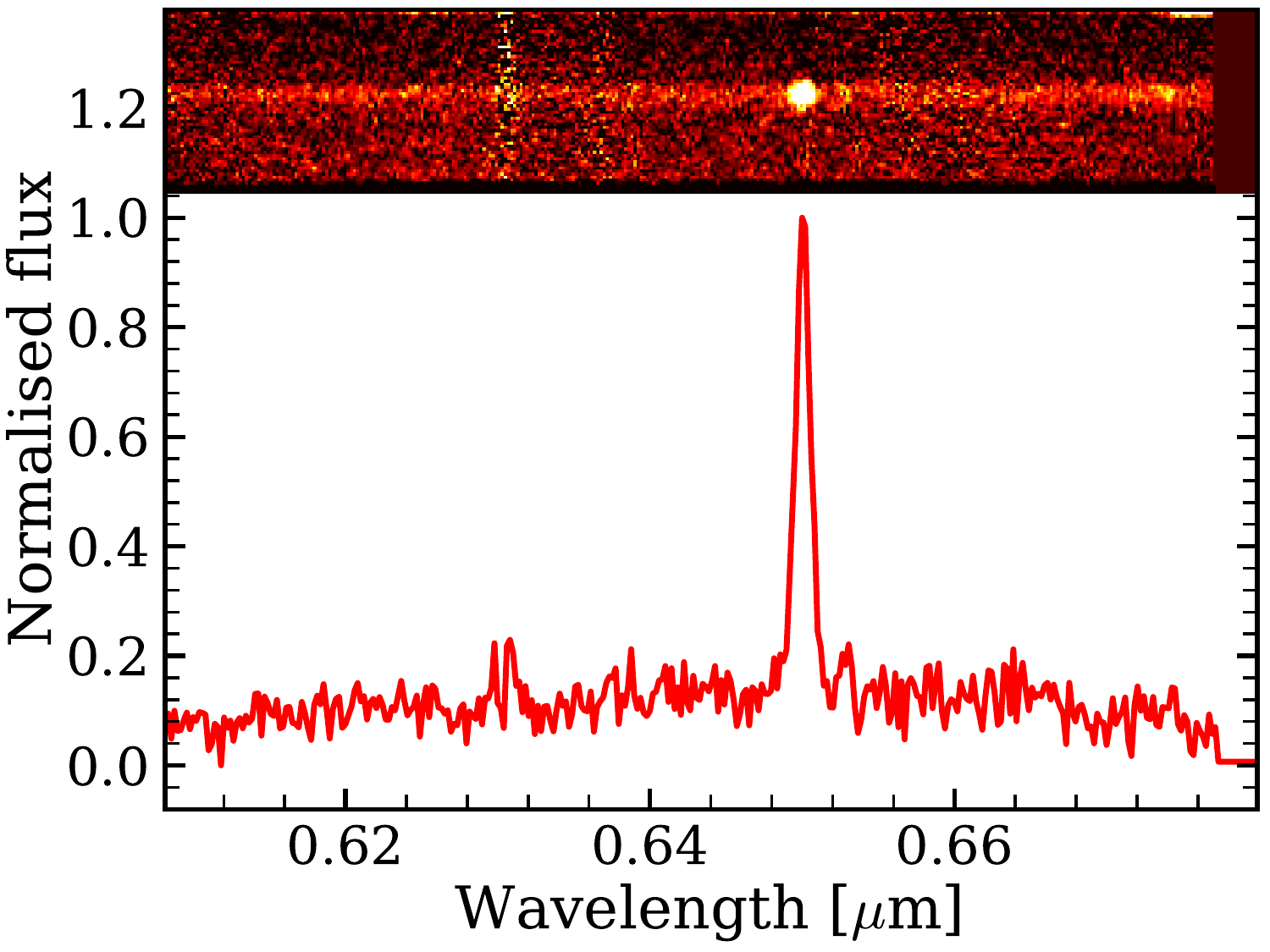}\\

	\includegraphics[width=0.3\textwidth]{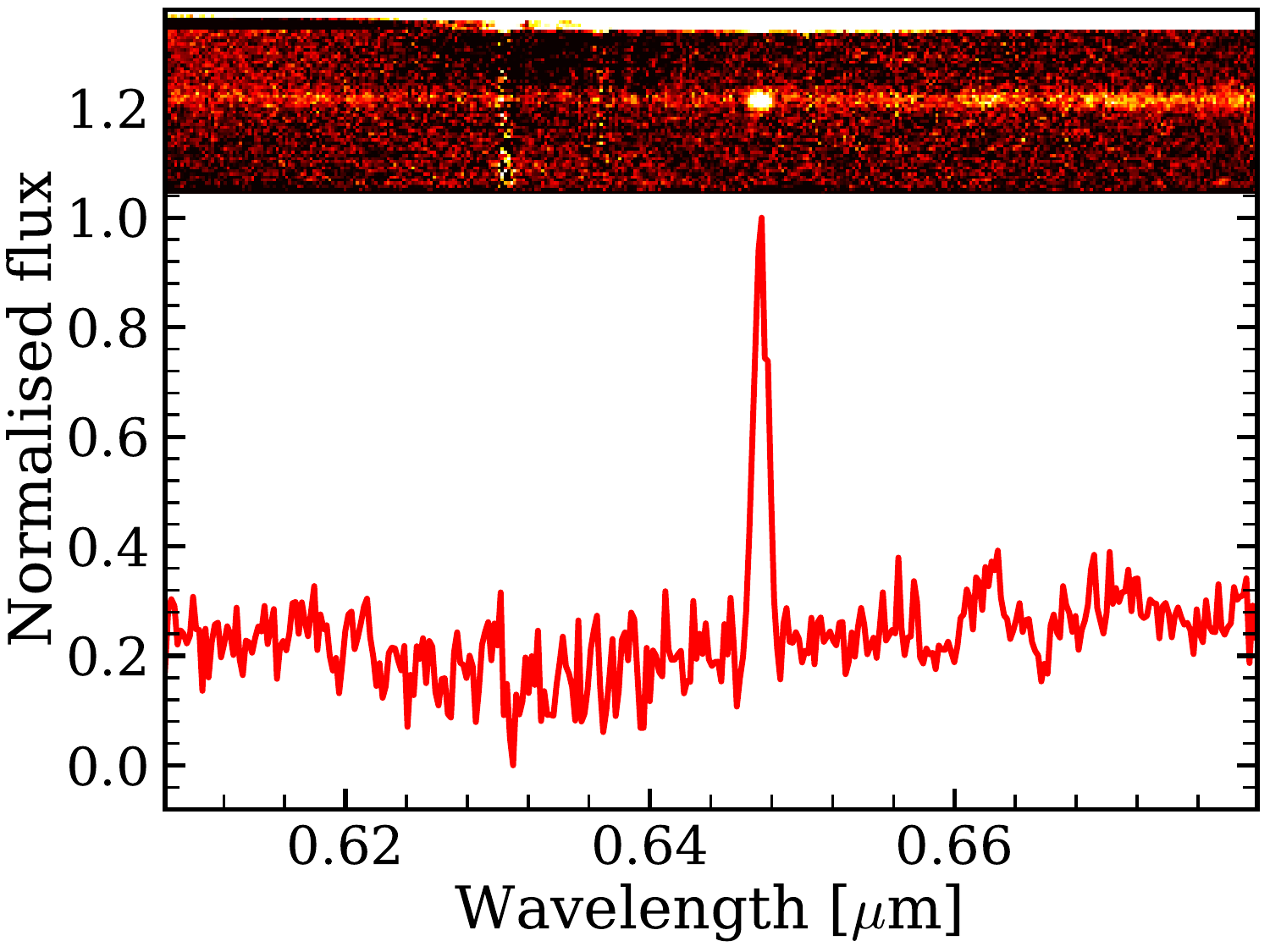}
	\includegraphics[width=0.3\textwidth]{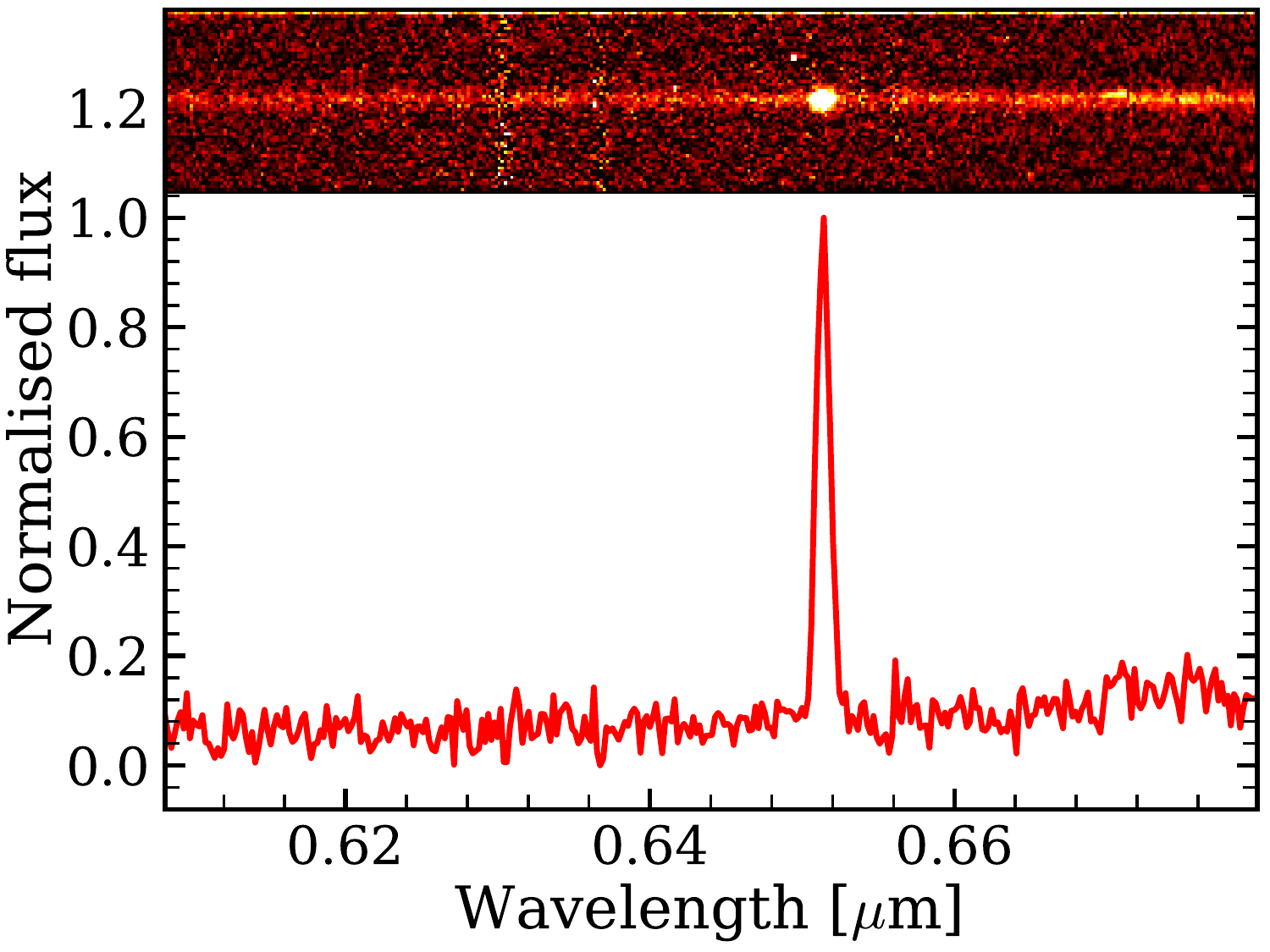}
	\includegraphics[width=0.3\textwidth]{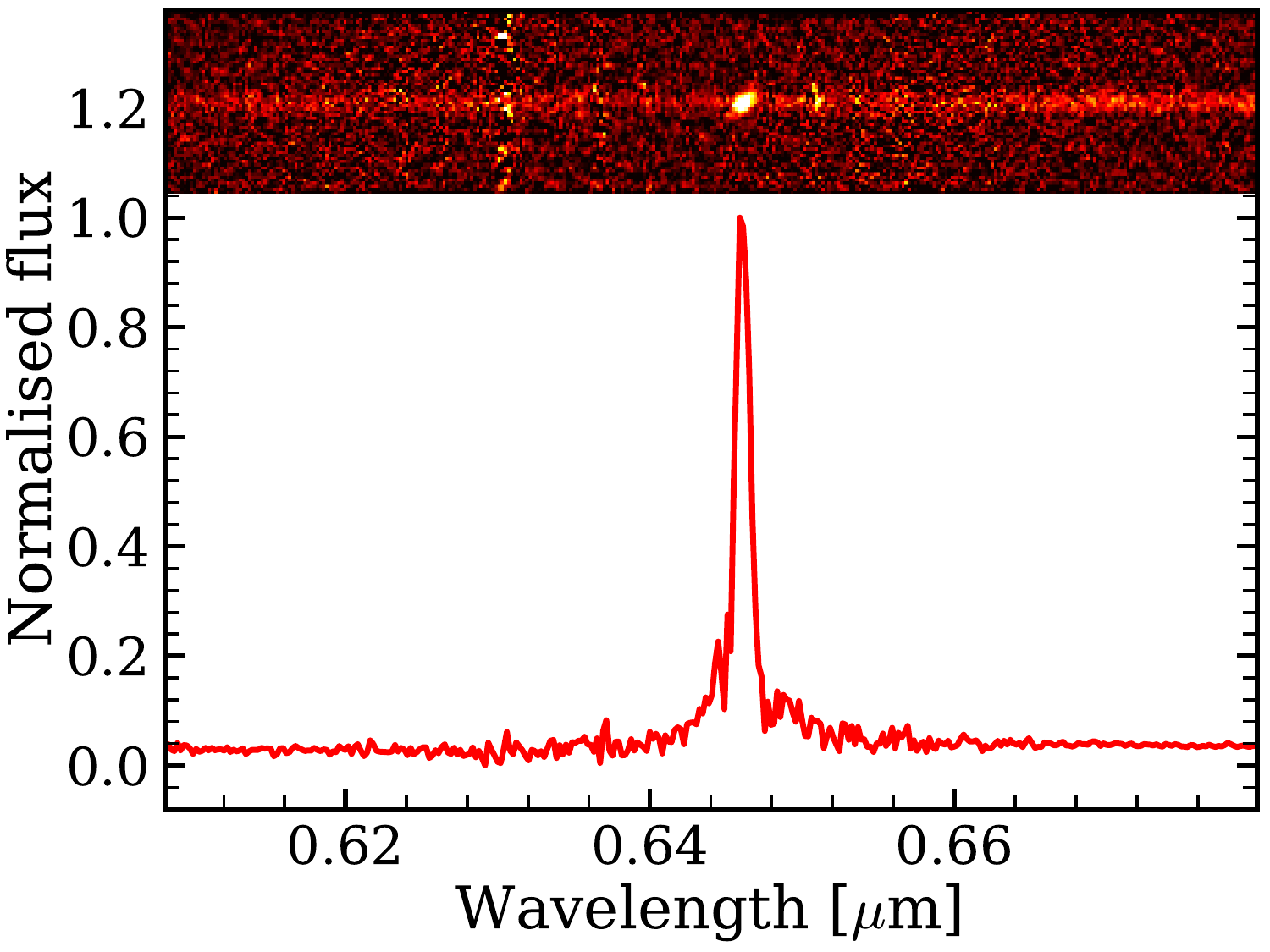} \\

	\includegraphics[width=0.3\textwidth]{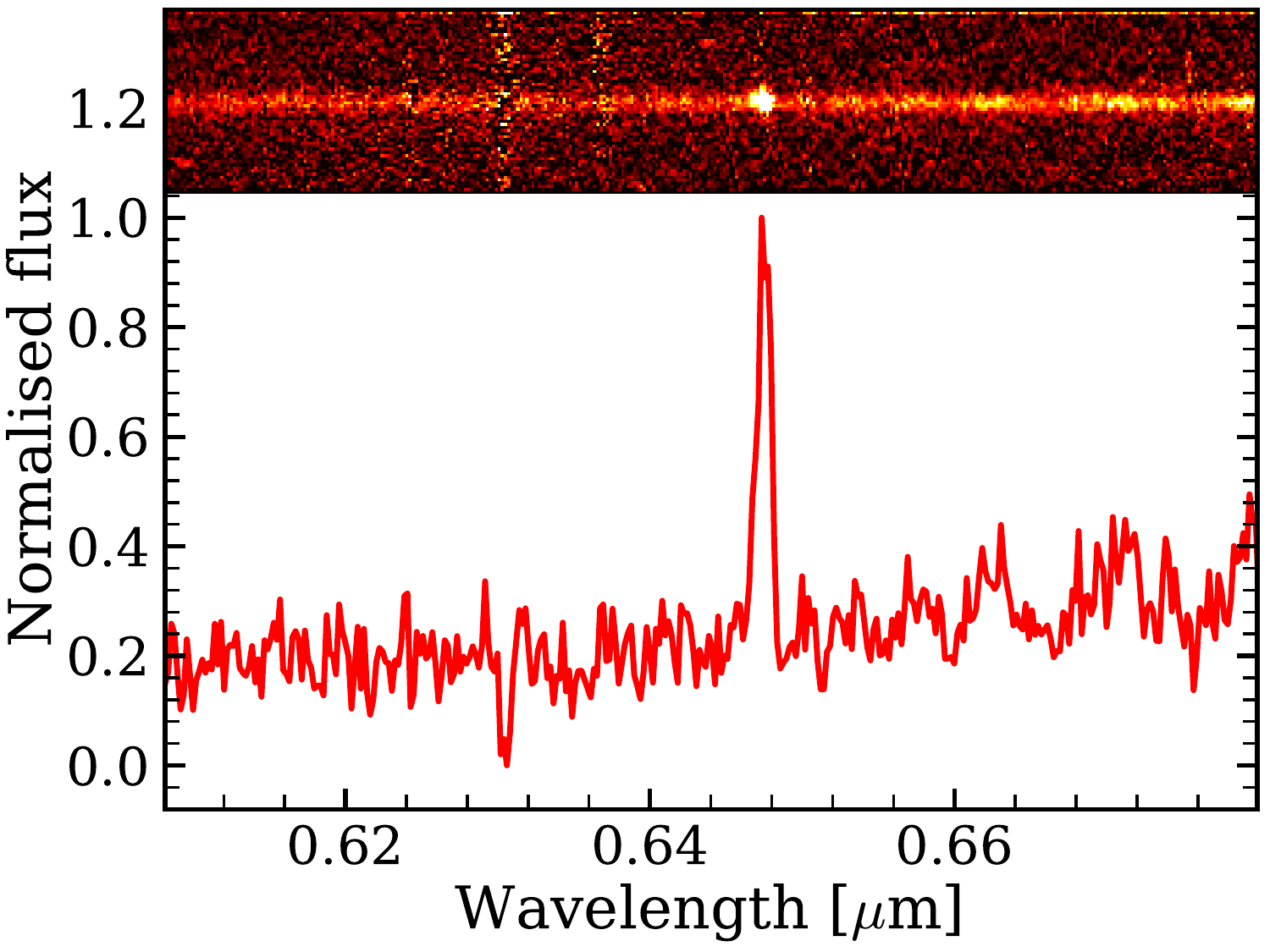}
	\includegraphics[width=0.3\textwidth]{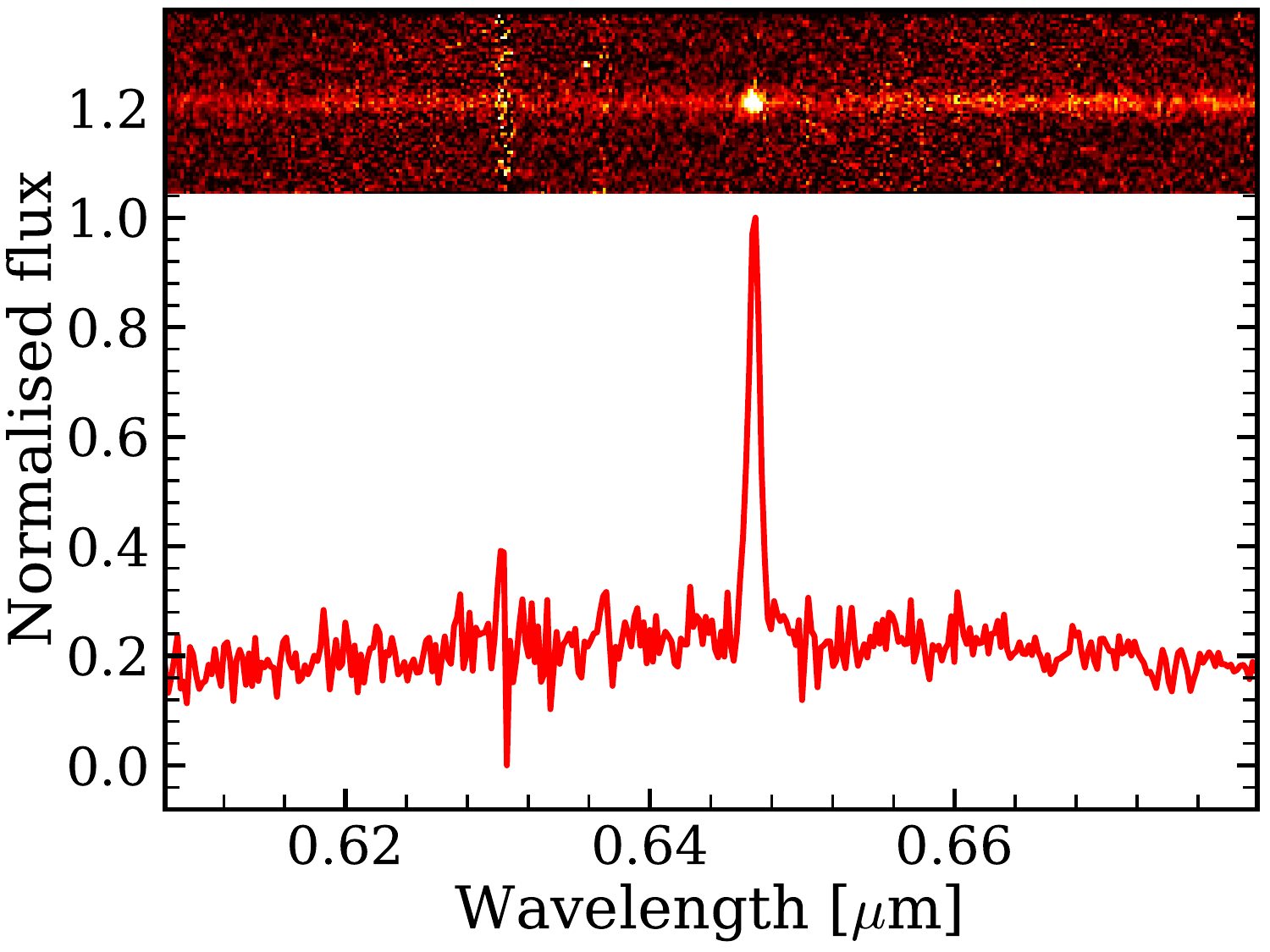}
	\includegraphics[width=0.3\textwidth]{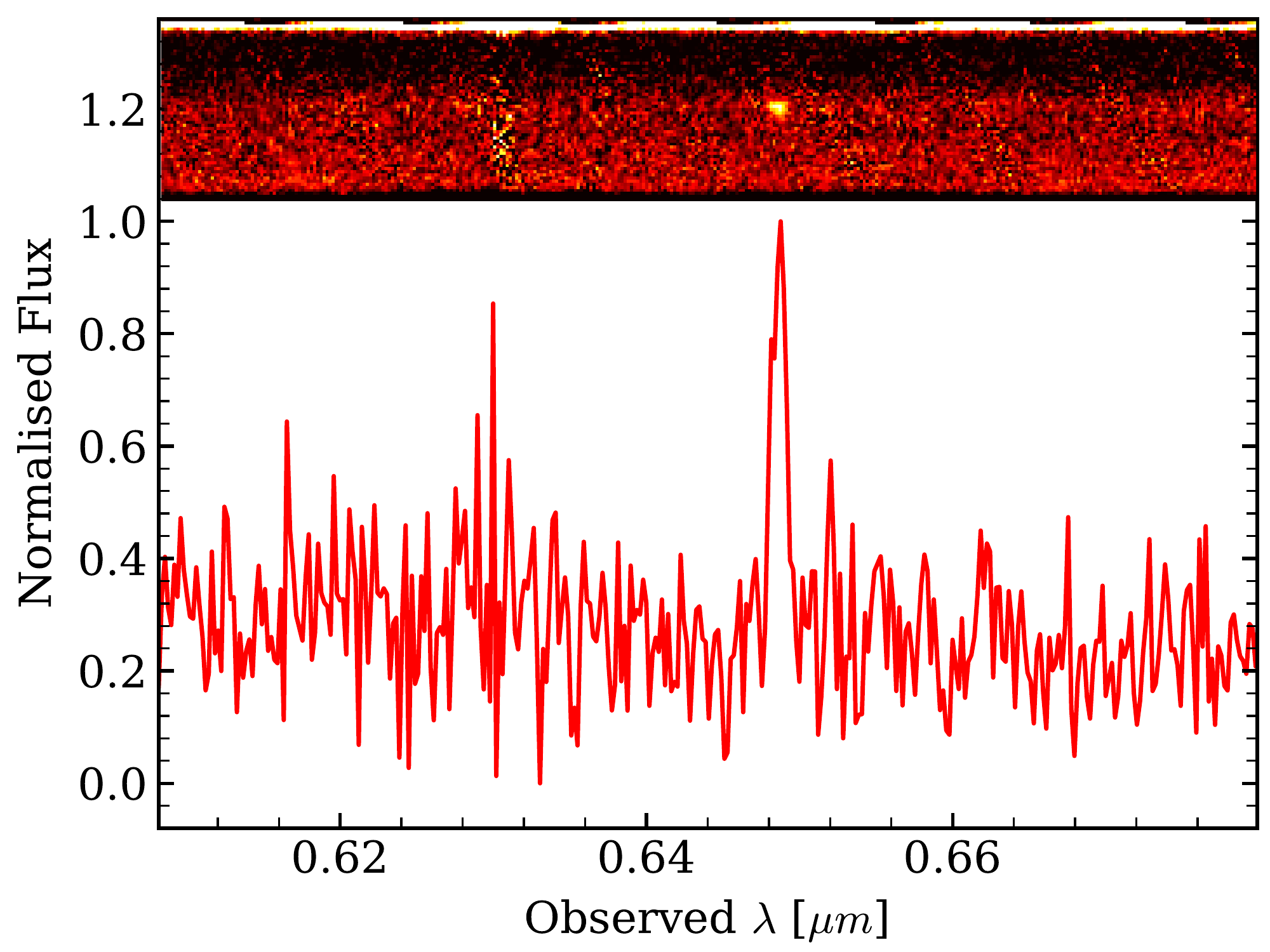}\\

	\includegraphics[width=0.3\textwidth]{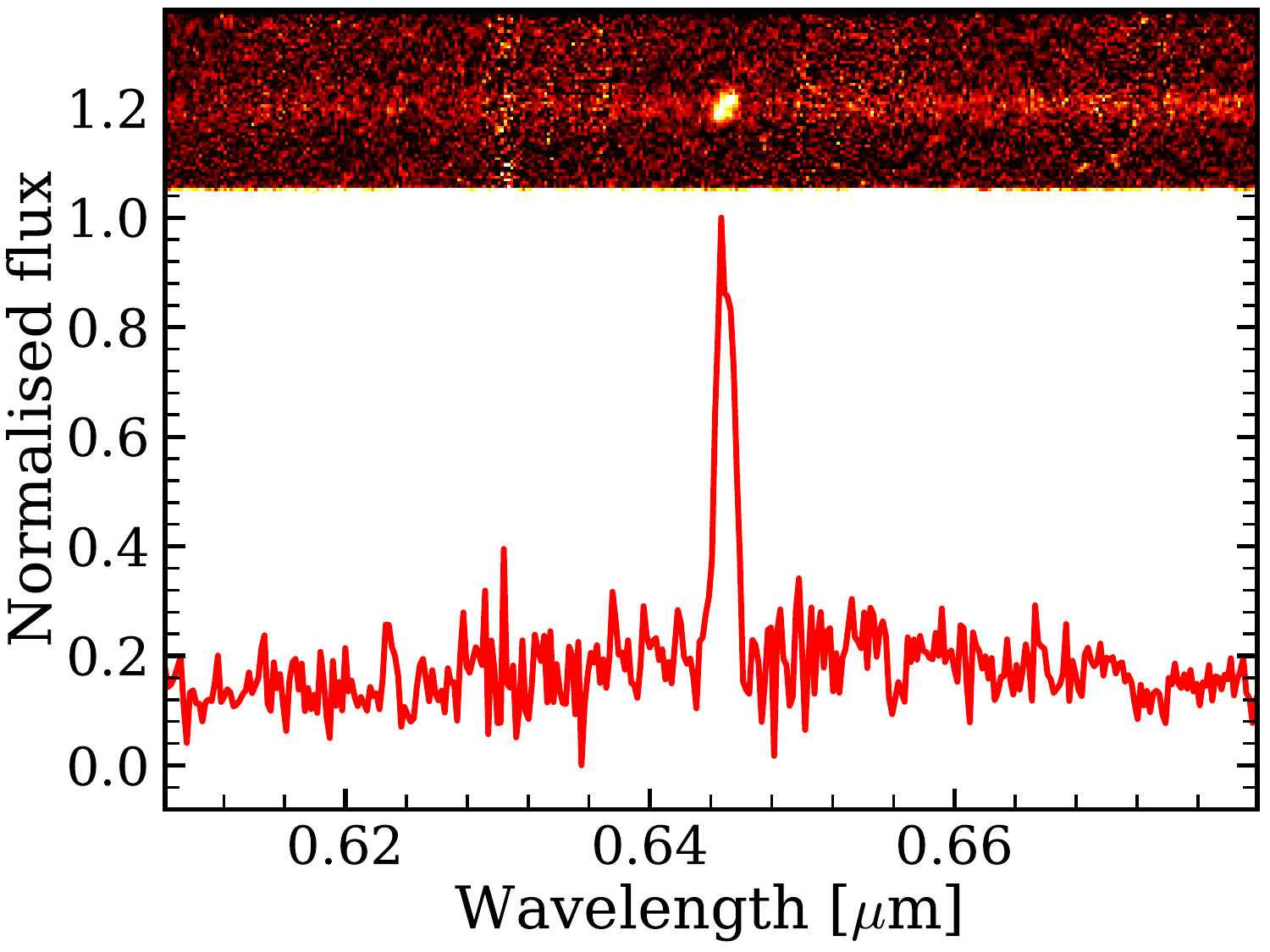}
	\includegraphics[width=0.3\textwidth]{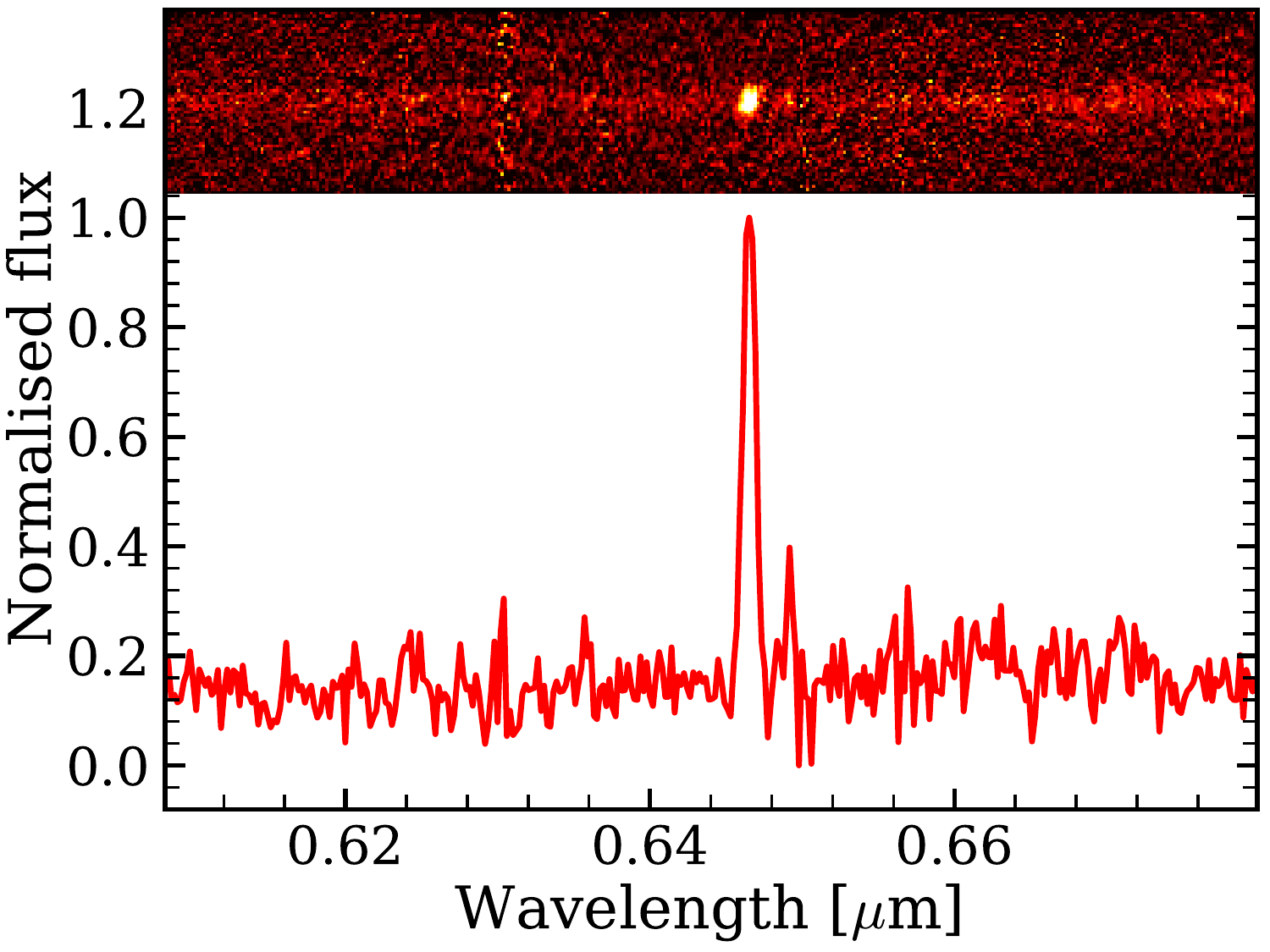}
	\includegraphics[width=0.3\textwidth]{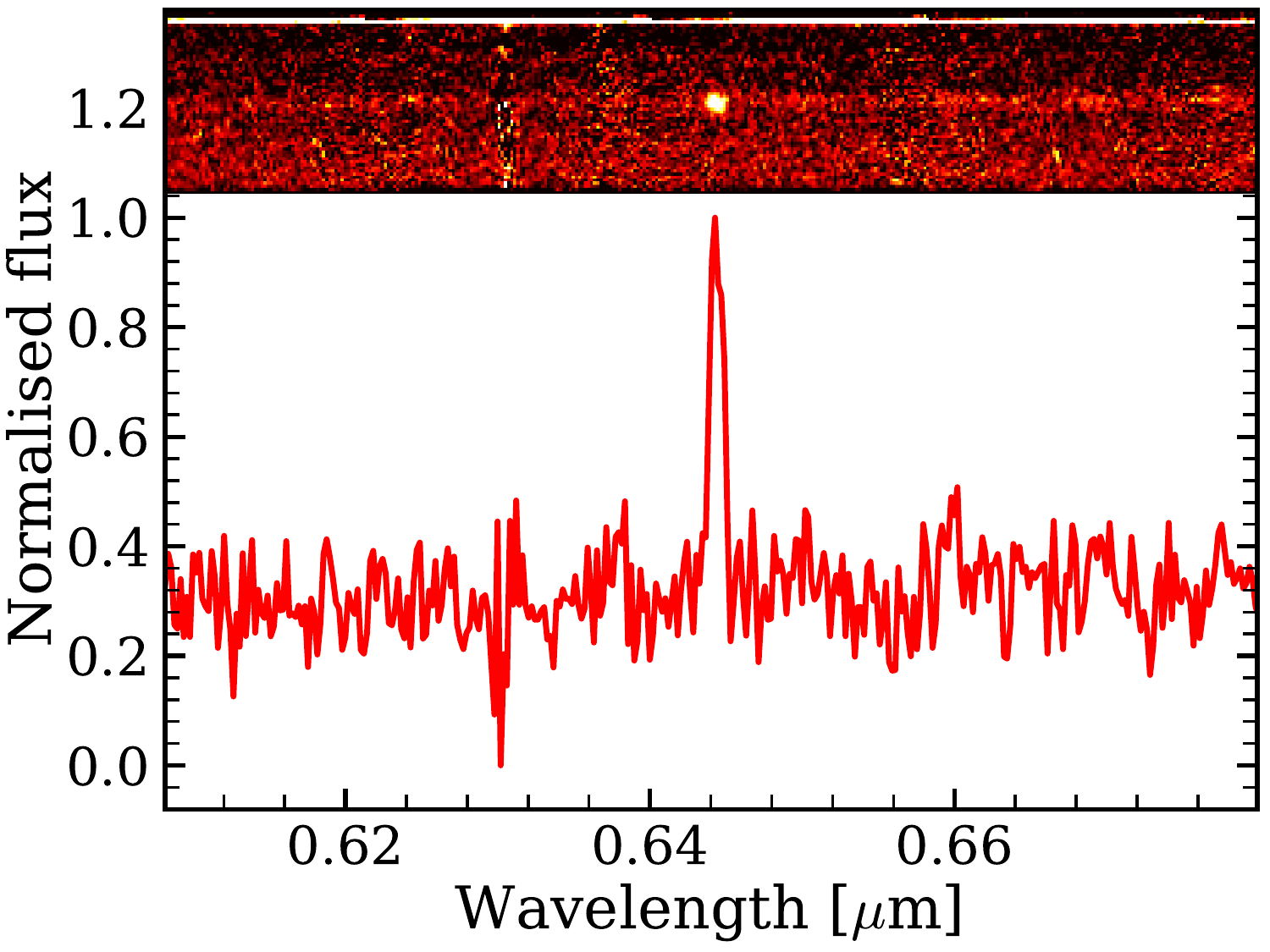}
        \caption{Example 1D and 2D spectrum}
         \label{fig:13}
\end{figure*}
}

\newpage

\begin{table*}
\scriptsize
\centering
\caption{The information of our observed galaxies.}
\label{tab:observation}
\begin{tabular}{ccccccc}
\hline
Slit & ID & R.~A. & Dec & R & phot-$z$ & spec-$z$ \\
      &      & (J2000) & (J2000) & (mag)& &  \\
\hline
S1-3   &   7448   &   03:32:36.02   &   $-$28:00:34.7   &   23.05   &   0.626   &   0.7348    \\
S1-5   &   3419   &   03:32:29.29   &   $-$28:02:27.6   &   23.18   &   0.751   &   0.6854    \\
S1-6   &   4783   &   03:32:28.68   &   $-$28:01:49.1   &   22.58   &   0.697  &   0.7356    \\
S1-10   &   1069   &   03:32:15.19   &   $-$28:03:30.3   &   23.06   &   0.695   &   0.7394    \\
S1-14   &   649   &   03:32:08.85   &   $-$28:03:43.0   &   21.53   &   0.716   &   0.7878    \\
S1-26   &   4398   &   03:32:02.62   &   $-$28:01:57.5   &   23.01   &   0.684   &   0.7380    \\
S1-38   &   12761   &   03:32:48.54   &   $-$27:58:04.3   &   22.78   &   0.757   &   0.7754    \\
S1-40   &   9700   &   03:32:35.93   &   $-$27:59:30.0   &   22.51   &   0.687   &   0.7351    \\
S1-42   &   13899   &   03:32:30.13   &   $-$27:57:30.8   &   22.52   &   0.715   &   0.6914    \\
S1-44   &   8587   &   03:32:45.35   &   $-$28:00:03.6   &   22.77   &   0.745   &   0.7813    \\
S1-52   &   9013   &   03:31:59.53   &   $-$27:59:49.9   &   22.78   &   0.698   &   0.7377    \\
S1-56   &   13664   &   03:31:59.96   &   $-$27:57:36.5   &   22.78   &   0.700   &   0.7362    \\
S1-64   &   11927   &   03:31:46.94   &   $-$27:58:26.4   &   22.52   &   0.749   &   0.7379    \\
S1-71   &   18594   &   03:33:04.98   &   $-$27:55:13.6   &   23.18   &   0.717   &   0.5803    \\
S1-77   &   16713   &   03:32:11.33   &   $-$27:56:09.9   &   22.40   &   0.693   &   0.7370    \\
S1-80   &   19335   &   03:32:08.85   &   $-$27:54:50.9   &   22.98   &   0.748   &   0.7361    \\
S1-86   &   19136   &   03:31:58.25   &   $-$27:54:56.6   &   22.10   &   0.728   &   0.7370    \\
S1-92   &   14520   &   03:32:04.95   &   $-$27:57:09.7   &   22.81   &   0.723   &   0.7997    \\
S1-93   &   17597   &   03:32:06.81   &   $-$27:55:41.1   &   23.12   &   0.697   &   0.7370    \\
S1-96   &   15029   &   03:31:52.39   &   $-$27:56:54.8   &   22.48   &   0.694   &   0.7432    \\
S1-102   &   15502   &   03:31:28.52   &   $-$27:56:40.3   &   22.91   &   0.674   &   0.7530    \\
S1-105   &   18547   &   03:31:27.68   &   $-$27:55:12.1   &   22.37   &   0.739   &   0.7475    \\
S1-106   &   19260   &   03:31:37.25   &   $-$27:54:50.4   &   22.94   &   0.673   &   0.7006    \\
S1-107   &   20182   &   03:31:36.34   &   $-$27:54:22.1   &   23.01   &   0.725   &   0.8139    \\
S1-108   &   22309   &   03:33:11.87   &   $-$27:53:23.6   &   22.50   &   0.715   &   0.7364    \\
S1-111   &   21175   &   03:33:10.90   &   $-$27:53:56.5   &   22.62   &   0.811   &   0.6893    \\
S1-113   &   23129   &   03:32:12.34   &   $-$27:52:57.2   &   22.83   &   0.661   &   0.7309    \\
S1-114   &   24760   &   03:32:11.80   &   $-$27:52:11.9   &   22.09   &   0.691   &   0.7305    \\
S1-117   &   27135   &   03:31:57.41   &   $-$27:50:59.7   &   22.46   &   0.735   &   0.8094    \\
S1-118   &   23479   &   03:31:42.72   &   $-$27:52:45.3   &   22.68   &   0.731   &   0.7357    \\
S1-120   &   24001   &   03:31:39.76   &   $-$27:52:32.0   &   22.97   &   0.689   &   0.6813    \\
S1-121   &   25346   &   03:31:34.13   &   $-$27:51:49.7   &   22.91   &   0.716   &   0.7846    \\
S1-122   &   26899   &   03:31:41.54   &   $-$27:51:07.1   &   22.64   &   0.711   &   0.7467    \\
S1-123   &   26915   &   03:33:17.67   &   $-$27:51:07.9   &   22.57   &   0.749   &   0.7070    \\
S1-124   &   24775   &   03:33:15.20   &   $-$27:52:11.3   &   21.71   &   0.711   &   0.6906    \\
S1-130   &   29085   &   03:31:54.04   &   $-$27:50:05.8   &   21.66   &   0.710   &   0.7426    \\
S1-131   &   31702   &   03:31:56.68   &   $-$27:48:46.8   &   22.96   &   0.719   &   0.7325    \\
S1-134   &   28807   &   03:31:49.65   &   $-$27:50:11.5   &   21.97   &   0.705   &   0.7362    \\
S1-144   &   37670   &   03:33:23.76   &   $-$27:45:56.4   &   22.11   &   0.735   &   0.6627    \\
S1-147   &   35612   &   03:33:09.46   &   $-$27:46:53.6   &   22.25   &   0.714   &   0.7339    \\
S1-149   &   38308   &   03:33:15.48   &   $-$27:45:36.7   &   23.12   &   0.717   &   0.8379    \\
S1-150   &   37758   &   03:32:59.90   &   $-$27:45:56.3   &   21.31   &   0.674   &   0.7338    \\
S1-157   &   34298   &   03:31:53.73   &   $-$27:47:27.6   &   22.71   &   0.717   &   0.7345    \\
S1-160   &   38348   &   03:31:47.68   &   $-$27:45:34.5   &   23.10   &   0.759   &   0.7357    \\
S1-161   &   38832   &   03:31:50.44   &   $-$27:45:21.4   &   22.26   &   0.678  &   0.7322    \\
S1-179   &   46040   &   03:32:44.65   &   $-$27:42:02.4   &   22.26   &   0.711   &   0.7090    \\
S1-181   &   47576   &   03:32:45.83   &   $-$27:41:17.1   &   22.94   &   0.727   &   0.7323    \\
S1-187   &   43498   &   03:31:57.94   &   $-$27:43:08.9   &   22.97   &   0.698   &   0.7294    \\
S1-188   &   46768   &   03:32:00.55   &   $-$27:41:43.1   &   22.03   &   0.725   &   0.7286    \\
S1-209   &   53041   &   03:32:40.93   &   $-$27:38:46.7   &   22.23   &   0.714   &   0.7366    \\
S1-210   &   53346   &   03:32:45.52   &   $-$27:38:40.2   &   21.84   &   0.715   &   0.7341    \\
S1-211   &   53492   &   03:32:33.55   &   $-$27:38:31.4   &   23.11   &   0.711   &   0.7330    \\
S1-212   &   52953   &   03:32:20.46   &   $-$27:38:47.2   &   22.83   &   0.675   &   0.7350    \\
S1-225   &   57449   &   03:32:28.19   &   $-$27:36:42.4   &   21.33   &   0.703   &   0.7333    \\
S1-232   &   42764   &   03:32:27.58   &   $-$27:43:31.4   &   22.92   &   0.685   &   0.7380    \\
S1-233   &   40914   &   03:32:23.16   &   $-$27:44:22.8   &   22.69   &   0.734   &   0.7373    \\
S1-250E   &   3139   &   03:32:11.85   &   $-$28:02:34.5   &   22.53   &   0.742   &   0.7017    \\
S1-285E   &   29193   &   03:31:36.90   &   $-$27:49:59.3   &   21.97   &   0.701   &   0.7167    \\
S1-291E   &   38339   &   03:31:45.90   &   $-$27:45:38.7   &   21.34   &   0.739   &   0.7372    \\
S1-292E   &   38874   &   03:31:48.47   &   $-$27:45:18.4   &   22.99   &   0.753   &   0.7375    \\
S1-314E   &   40013   &   03:32:20.67   &   $-$27:44:46.3   &   22.81   &   0.737   &   0.7252    \\
S2-9   &   7119   &   03:32:45.07   &   $-$28:00:42.6   &   23.39   &   0.716   &   0.7329    \\
S2-16   &   827   &   03:32:13.54   &   $-$28:03:36.1   &   23.39   &   0.666   &   0.7095    \\
S2-44   &   13095   &   03:32:44.67   &   $-$27:57:52.1   &   23.86   &   0.711   &   0.7734    \\
S2-63   &   16957   &   03:32:55.95   &   $-$27:56:00.5   &   23.88   &   0.730   &   0.7748    \\
S2-78   &   22528   &   03:32:04.35   &   $-$27:53:14.2   &   23.45   &   1.285   &   0.7441    \\
S2-100   &   30432   &   03:31:46.83   &   $-$27:49:22.8   &   23.94   &   -99.0   &   0.7353    \\
S2-125   &   34954   &   03:31:42.76   &   $-$27:47:07.3   &   23.36   &   0.808   &   0.7337    \\
S2-159   &   48680   &   03:32:55.30   &   $-$27:40:46.9   &   23.84   &   0.692   &   0.7336    \\
S2-187   &   29357   &   03:32:54.45   &   $-$27:49:55.5   &   23.90   &   -99.0   &   0.7327    \\
S2-198   &   50082   &   03:32:28.98   &   $-$27:40:06.2   &   23.90   &   0.616   &   0.7324    \\
S2-216   &   24627   &   03:32:18.57   &   $-$27:52:12.8   &   23.43   &   0.805   &   0.7964    \\
S2-217   &   29835   &   03:32:22.09   &   $-$27:49:39.9   &   23.74   &   1.103   &   0.7306    \\
S2-241E   &   12150   &   03:32:05.23   &   $-$27:58:18.2   &   23.47   &   0.604   &   0.7279    \\
S2-266E   &   54514   &   03:32:38.59   &   $-$27:38:03.3   &   23.88   &   0.754   &   0.7505    \\

\hline
\end{tabular}
\end{table*}


\bsp	
\label{lastpage}
\end{document}